\documentclass[graybox]{svmult}

\usepackage{type1cm}        % activate if the above 3 fonts are
\usepackage{makeidx}         % allows index generation
\usepackage{graphicx}        % standard LaTeX graphics tool
\usepackage{multicol}        % used for the two-column index
\usepackage[bottom]{footmisc}% places footnotes at page bottom

\usepackage{newtxtext}       % 
\usepackage{newtxmath}       % selects Times Roman as basic font

\makeindex             % used for the subject index
\usepackage{bm}
\usepackage{physics}
\usepackage{subcaption}

\DeclareMathAlphabet{\mathsfit}{T1}{\sfdefault}{\mddefault}{\sldefault}
\SetMathAlphabet{\mathsfit}{bold}{T1}{\sfdefault}{\bfdefault}{\sldefault}
\newcommand{\ten}[1]{\bm{\mathsfit{#1}}}

\newcommand{\ns}{n_0} % n_\text{s}
\newcommand{\kt}{k_\text{B}T}
\newcommand{\eps}{\varepsilon}
\newcommand{\lB}{\ell_\text{B}}
\newcommand{\debye}{\lambda_\text{D}}
\newcommand{\lGC}{\ell_\text{GC}}
\newcommand{\lDu}{\ell_\text{Du}}
\newcommand{\rhoe}{\rho_\text{e}}
\newcommand{\Vs}{V_\text{s}}
\newcommand{\vs}{v_\text{s}}
\newcommand{\zs}{z_\text{s}}
\newcommand{\phis}{\phi_\text{s}}

\newcommand{\Rch}{R_\text{ch}}
\newcommand{\Zch}{Z_\text{ch}}
\newcommand{\Rl}{R_\text{L}}
\newcommand{\Sstr}{S_\text{str}}
\newcommand{\Sdoc}{S_\text{doc}}
\newcommand{\sgn}{\text{sgn}}
\newcommand{\tosm}{\text{to}}
\newcommand{\tel}{\text{te}}

\newcommand{\diff}[2]{\frac{\mathrm{d}#1}{\mathrm{d}#2}}
\newcommand{\ddiff}[2]{\frac{\mathrm{d}^2 #1}{\mathrm{d}#2^2}}
\newcommand{\pdiff}[2]{\frac{\partial #1}{\partial #2}}

\newcommand{\ilm}{Univ Lyon, Univ Claude Bernard Lyon 1, CNRS, Institut Lumi\`ere Mati\`ere, F-69622, Villeurbanne, France}

\begin{document}

\title*{Energy conversion at water-solid interfaces using electrokinetic effects}
\author{Cecilia Herrero, Aymeric Allemand, Samy Merabia, Anne-Laure Biance, and Laurent Joly}
\authorrunning{C. Herrero, A. Allemand, S. Merabia, A.-L. Biance, and L. Joly}
\institute{Cecilia Herrero \at \ilm
%, \email{name@email.address}
\and 
Aymeric Allemand \at \ilm
%, \email{name@email.address}
\and 
Samy Merabia \at \ilm
%, \email{name@email.address}
\and 
Anne-Laure Biance \at \ilm
%, \email{name@email.address}
\and Laurent Joly \at \ilm, \email{laurent.joly@univ-lyon1.fr}
}
\maketitle

\abstract*{Our Society is in high need of alternatives to fossil fuels. Nanoporous systems filled with aqueous electrolytes show great promises for harvesting the osmotic energy of sea water or waste heat. At the core of energy conversion in such nanofluidic systems lie the so-called electrokinetic effects, coupling thermodynamic gradients and fluxes of different types (hydrodynamical, electrical, chemical, thermal) at electrified water-solid interfaces. This chapter starts by introducing the framework of linear irreversible thermodynamics, and how the latter can be used to describe the direct and coupled responses of a fluidic system, providing general relations between the different response coefficients. The chapter then focuses on the so-called osmotic flows, generated by non-hydrodynamic actuation at liquid-solid interfaces, and illustrate how the induced fluxes can be related to the microscopic properties of the water-solid interface. Finally, the chapter moves to electricity production from non-electric actuation, and discusses in particular the performance of nanofluidic systems for the harvesting of osmotic energy and waste heat.}

\abstract{Our Society is in high need of alternatives to fossil fuels. Nanoporous systems filled with aqueous electrolytes show great promises for harvesting the osmotic energy of sea water or waste heat. At the core of energy conversion in such nanofluidic systems lie the so-called electrokinetic effects, coupling thermodynamic gradients and fluxes of different types (hydrodynamical, electrical, chemical, thermal) at electrified water-solid interfaces. This chapter starts by introducing the framework of linear irreversible thermodynamics, and how the latter can be used to describe the direct and coupled responses of a fluidic system, providing general relations between the different response coefficients. The chapter then focuses on the so-called osmotic flows, generated by non-hydrodynamic actuation at liquid-solid interfaces, and illustrate how the induced fluxes can be related to the microscopic properties of the water-solid interface. Finally, the chapter moves to electricity production from non-electric actuation, and discusses in particular the performance of nanofluidic systems for the harvesting of osmotic energy and waste heat.}

%\tableofcontents

\section{Introduction: nanofluidic systems for energy conversion} 

Climate change continues progressing at an alarming rate. Among the causes of human impact on this evolution, the use of fossil fuels appears critical and some new methods of renewable  energy extractions are largely explored \cite{Dale2019bp}. Among them, taking advantage of \emph{blue energy} generated by water salinity gradients \cite{Pattle1954}, 
or \emph{waste heat} \cite{Straub2016}, inducing irreversibly lost temperature gradients, appears promising \cite{Elimelech2011}. Some specific systems, nanoporous membranes filled with a liquid solution, are good candidates for these processes. Indeed, living cell machinery already uses these complex processes to selectively transport ions and solutes in subnanometric proteinic channels  \cite{Greger1996}, paving the way for new nanofluidic system designs in mimicking nature \cite{Marbach2016}. 
In the following, we will detail, from a fundamental point of view, what are the prerequisites for energy conversion in nanofluidic systems. In a first part, we will detail how coupled transport properties can emerge  at water-solid interfaces. We will then describe how external gradients (salinity, temperature, electrostatic potential gradients) can induce so-called \textit{osmotic} flows near the interface. Finally, in a last part, in the context of useful energy harvesting, we will show how electricity can also be produced.

\section{Electrokinetic effects: coupled transport at water-solid interfaces} 

Electrokinetic (EK) effects refer to the coupled response of a fluidic system, where an actuation of a certain type induces a flux of another type. The most well-known EK effects are electro-osmosis (the flow induced by an electric field)  \cite{reuss1809charge} and streaming current (the electric current induced by a pressure gradient), but all types of transport (hydrodynamical, electrical, chemical, thermal) are coupled through EK effects. In this section, we will first briefly introduce the framework of linear irreversible thermodynamics, which can be used to describe the direct and coupled responses of a fluidic system, and can provide general relations between the different response coefficients; we will then discuss the physical origins of coupled transport in fluidic systems.

\subsection{Direct and coupled response of a fluidic system: general properties} \label{sec:irreversible_thermodynamics}

When a system is driven weakly out of equilibrium by small thermodynamic gradients, one can develop the resulting fluxes as a Taylor series of the gradients. Linear response theory provides a theoretical background to understand the first order in this development, i.e. the linear relation between applied gradients and resulting fluxes close to equilibrium. 
This linear relation is fully encoded in a response matrix, and this is our aim here to present this matrix and its properties, which can be derived using the tools of statistical physics. 
For further reading, complementary discussions can be found in Refs.~\cite{deGrootMazur,balian2006,kjelstrup2008non,kjelstrup2017,livi2017}.

\subsubsection{Definitions}

\paragraph*{\textbf{Affinities or generalized forces}}

First, one can consider two coupled systems, denoted $a$ and $b$ in the following. The thermodynamic state of these two subsystems is characterized by the knowledge of a set of extensive variables $X_i^{(a)}$ and $X_i^{(b)}$. For the fluidic systems of concern here -- considering a fluid mixture of different particles of type $n$, the set of $X_i^{(a)}$ includes the volume $\mathcal{V}$ of the subsystem $a$, its internal energy $\mathcal{U}$, the total number of particles $N$, the number $N_n$ of particles of type $n$, and the electric charge $Q$. If the two subsystems are separated by a membrane that allows these extensive variables to be transferred from one system to the other, each variable $X_i^{(a)}$ may vary in time so that $X_i^{(a)}+X_i^{(b)}$ is constant (the \emph{total} system is assumed to be closed). 
At equilibrium, the entropy of the total system should obey:
\begin{equation}
\label{eq:variation_S}
    \dd S=\sum_i \left(\frac{\partial S}{\partial X_i^{(a)}}-\frac{\partial S}{\partial X_i^{(b)}}\right) \dd X_i^{(a)}=0. 
\end{equation}
As this latter equation should hold whatever the variation $dX_i^{(a)}$, we have, at equilibrium:
\begin{equation}\label{eq:definition_affinity}
    \mathcal{F}_i=\frac{\partial S}{\partial X_i^{(a)}}-\frac{\partial S}{\partial X_i^{(b)}}=0 , 
\end{equation}
where we have defined the affinities (or the generalized forces) $\mathcal{F}_i$. 
The affinities of an isotropic fluidic system may be calculated using the Gibbs equality:
\begin{equation}
\label{eq:Gibbs_identity}
    \dd S=\frac{1}{T} \dd \mathcal{U}+ \frac{P}{T} \dd \mathcal{V} - \sum_{n} \frac{\mu_n}{T} \dd N_{n} + \frac{V}{T} \dd Q , 
\end{equation}
where $T$ is the temperature, $P$ the pressure, $\mu_n$ the chemical potential of the constituent $n$ and $V$
the electrostatic potential. The affinities corresponding to the internal energy $\mathcal{U}$, the volume $\mathcal{V}$, the number of particles of type $n$ and the charge are given respectively by: 
\begin{equation}
    \label{eq:affinities}
    \frac{\partial S}{\partial \mathcal{U}}=\frac{1}{T} \quad  \frac{\partial S}{\partial \mathcal{V}}=\frac{P}{T} \quad \frac{\partial S}{\partial N_{n}}=-\frac{\mu_{n}}{T}\quad \mathrm{and} \quad \frac{\partial S}{\partial Q}=\frac{V}{T} . 
\end{equation}
The affinities may be hence calculated using Eq.~\eqref{eq:definition_affinity}. For instance, the affinity corresponding to the internal energy $\mathcal U$ is:
\begin{equation}
    \label{eq:affinity_internal_energy}
    \mathcal F_{\mathcal U} = \frac{1}{T^{(a)}}-\frac{1}{T^{(b)}} . 
\end{equation}
So far, we have considered a system at rest in the absence of macroscopic motion. Given the importance of hydrodynamic flows in the following, we should discuss the affinities associated to the fluid momenta $p_{\alpha}=m v_{\alpha}$, where $m$ is the mass of the fluid molecules and $v_{\alpha}$ the components of the fluid velocity with $\alpha \in {x,y,z}$. To proceed, let us consider a fluid system driven out of equilibrium by a pressure gradient which induces fluid motion. In the frame where the fluid system is at rest, its entropy is $S(\mathcal U,\mathcal V,N_{n},Q)$. In the frame where the system has a velocity $\vec v$, its entropy is $S(\mathcal U+ \sum_\alpha p_\alpha^2/(2m),\mathcal V,N_{n},Q)$ where we have expressed the fact that the entropy is \emph{Galilean invariant}, i.e. it does not depend on the choice of the frame of reference. These considerations allow us to define the affinity corresponding to the system momenta through:
\begin{equation}
    \label{eq:affinities_momenta}
     \frac{\partial S}{\partial p_{\alpha}}= - \frac{p_{\alpha}}{m} \left (\frac{\partial S}{\partial \mathcal U} \right)_{\mathcal V,N_{n},Q} = -\frac {v_{\alpha}}{T} \quad \alpha \in {x,y,z} . 
\end{equation}
At this stage, it should be remarked that the affinity associated to the fluid momenta, Eq.~\eqref{eq:affinities_momenta}, is a \emph{vectorial} quantity, and from this respect has a particular character as compared to the other affinities defined in Eq.~\eqref{eq:affinities}. 
This will be of special concern when we will discuss the consequences of the symmetry of the system on its out-of-equilibrium response.

\paragraph*{\textbf{Fluxes and Onsager matrix}}

Now, consider a situation where the system is driven out-of-equilibrium through some constraints imposed on the total system.
These constraints may be, for instance, the coupling with thermostats at different temperatures, with a reservoir of particles, 
or with an open-circuit. In that case, the variables $X_i^{(a)}$ may be transported from one subsystem to the other. Accordingly, the membrane separating the systems $(a)$ and $(b)$ will be crossed by a flux $J_i$ defined by:
\begin{equation}
    \label{eq:definition_flux}
    J_i^{(a)} = \frac{\partial X_i^{(a)}}{\partial t} , 
\end{equation}
and, since $X_i$ is a conserved variable one should have: $J_i^{(a)}=-J_i^{(b)}$. 
At equilibrium, all the fluxes vanish. If the system is driven not too far from equilibrium, one can perform a development at the first order of the flux with respect to the affinities (which vanish at equilibrium):
\begin{equation}
    \label{eq:relation_flux_affinities}
    J_i^{(a)}=\sum_j \mathcal{L}_{ij} {\mathcal F}_j , \quad \text{where} \quad \mathcal{L}_{ij}=\left(\frac{\partial J_i}{{\mathcal \partial {\mathcal F}}_j}\right)_{{{\mathcal F}_{k\neq j}}=0} . 
\end{equation}
The phenomenological relations in  Eq.~\eqref{eq:relation_flux_affinities} define the 
elements of the Onsager matrix $\mathcal{L}_{ij}$. Using the Onsager regression principle \cite{onsager1931_1,onsager1931_2,casimir1945,livi2017,Brunet2004}, one can show that the matrix elements $\mathcal{L}_{ij}$ are symmetric:
\begin{equation}
    \label{eq:symmetry_Onsager}
\mathcal{L}_{ij}=\mathcal{L}_{ji} . 
\end{equation}
This latter relation holds as the extensive variables of concern here are invariant under time reversal. Another important property of the coefficients $\mathcal{L}_{ij}$ may be obtained by considering the {total} entropy production {rate}:
\begin{equation}
\label{eq:bilinear_form_entropy}
    {\frac{\dd_\text{i} S}{\dd t}} = \sum_{i,j} \mathcal{L}_{ij} {\mathcal F}_i {\mathcal F}_j \ge 0 , 
\end{equation}
which turns out to be a bilinear function of the generalized forces. Because the total system (including the thermostat) is isolated, the second principle of thermodynamics states that the entropy production rate should be positive, \textit{i.e.}, ${\dd_\text{i} S/\dd t} \ge 0$, and should vanish at equilibrium, consistently with the fact that the generalized forces are null. The positiveness of the entropy production holds for any {\em small} affinity ${\mathcal F}_i$. Therefore, 
$\mathcal{L}_{ii} {\mathcal F}_i^2 \ge 0$, yielding the equality:
\begin{equation}
    \mathcal{L}_{ii} > 0 . 
\end{equation}
The positive character of the total entropy production also implies that the Onsager matrix -- which was introduced in Eq.~\eqref{eq:relation_flux_affinities} -- is positive definite. 

Another important practical consequence of the bilinear form, Eq.~\eqref{eq:bilinear_form_entropy}, may be used to identify the fluxes if we know the form of the entropy production ${\dd_\text{i} S/\dd t}$: 
\begin{equation}
    \label{eq:form_bilinear_identification_fluxes}
    {\frac{\dd_\text{i} S}{\dd t}} = \sum_i J_i {\mathcal F}_i , 
\end{equation}
with $J_i = \sum_j \mathcal {L}_{ij} {\mathcal F}_j$. 
The identification of the couples fluxes/forces may be useful in practice in situations for which we can evaluate simply the entropy production ${\dd_\text{i} S/\dd t}$. We will show in the following an example of that approach when we will deal with the response of nanofluidic devices.

\paragraph*{\textbf{Local quantities}}

We now generalize the previous considerations to the case of a continuous system, whose thermodynamic state may vary continuously in space under the action of constraints imposed to the system. To describe continuous systems, it is tempting to define mesoscopic subsystems $(a)$ and use locally thermodynamic identities such as Eq.~\eqref{eq:Gibbs_identity}. This is, however, only possible under the assumption of \emph{local thermodynamic equilibrium}. Under this assumption, one may define local 
{specific quantities $x_i(\vec r,t)$ (i.e., quantities per unit mass),} so that the value of the variable $X_i$ for the total system writes:
\begin{equation}
X_i(t) = \int_{\mathcal V} {\rho(\vec r,t)}\, x_i(\vec r,t) \,\dd  \vec r ,   
\end{equation}
where $\mathcal V$ is the volume of the whole system{, and $\rho(\vec r,t)$ the density}. This decomposition is valid if there exists a separation of time scales between microscopic processes and macroscopic ones. For condensed fluidic systems, the time scale controlling the relaxation to local thermodynamic equilibrium is given by the typical collision time between molecules, which is on the order of $1$\,ps. The assumption of local equilibrium also implies that the subsystems that we consider have mesoscopic length scales. Quantitatively, this means that every subsystem contains enough particles so that it behaves as a thermodynamic system, whose state is characterized by a small number of thermodynamic variables \cite{pottier2009}. 

Under these conditions, the variation of the {local specific} entropy writes:
$\dd s = \sum_i \left(\frac{\partial s}{\partial x_i}\right)_{x_{j \neq i}} \dd x_i$, and it is natural to define the local affinities 
by:
\begin{equation}
    \label{eq:affinities_continuous_system}
    \vec f_i(\vec r,t) = {-} \vec \nabla \left(\frac{\partial s}{\partial x_i}\right)_{x_{j \neq i}} . 
\end{equation}
Correspondingly, the fluxes $\vec j_i$ corresponding to the conserved variables {$\rho x_i$} may be defined based on the local conservation equation:
\begin{equation}
\label{eq:local_conservation}
\frac{\partial {(\rho x_i)}}{\partial t} = - \nabla \cdot \vec j_i . 
\end{equation}
The phenomenological relation between the fluxes $\vec j_i$ and the affinities then becomes:
\begin{equation}
 \label{eq:relation_flux_affinities_continuous}
    \vec j_i(\vec r,t) = \sum_j L_{ij} \vec f_j(\vec r,t) , 
\end{equation}
where $L_{ij}$ are the local Onsager coefficients. Note that, in general, $L_{ij}$ is a tensor so that Eq.~\eqref{eq:relation_flux_affinities_continuous} reads:
\begin{equation}
    \label{eq:relation_flux_affinities_vectorial}
    j_{i,\alpha}(\vec r,t) = \sum_j L_{ij,\alpha\beta} f_{j,\beta}(\vec r,t)
\end{equation}
where Einstein summation convention has been used. However, in the common case of isotropic media, the tensor $L_{ij}$ is diagonal and proportional to the identity tensor~:~$L_{ij,\alpha\beta}=L_{ij}\delta_{\alpha\beta}$. 

Equation~\eqref{eq:relation_flux_affinities_continuous} implies that the response of the system is instantaneous and local in space. This is a result of the existence of a separation of time scales between microscopic and macroscopic processes and the existence of local processes ensuring the relaxation to equilibrium. In rarefied media, the relation between fluxes and affinities may be not instantaneous and not local in space, as relaxation towards equilibrium occurs through infrequent collisions between molecules travelling ballistically. 

For some specific variables, the flux $\vec j_i$ may be non vanishing in the equilibrium state in the absence of gradients of the considered intensive quantity, and the relation between the flux and the affinities should read:
\begin{equation}
 \delta \vec j_i(\vec r,t) = \sum_j L_{ij} \vec f_j(\vec r,t)
\end{equation}
where $\delta \vec j_i=\vec j_i-\vec j_i^{\rm eq}$ is the deviation to the \emph{equilibrium} flux $\vec j_i^{\rm eq}$. As an example we will see soon, one needs to subtract an equilibrium flux when treating hydrodynamic flows.

{Finally, within the local formulation, the density of entropy production writes: 
\begin{equation}
\label{eq:bilinear_form_entropy_local}
    \sigma_s=\sum_{i,j} L_{ij} \vec f_i \cdot \vec f_j = \sum_i \delta \vec j_i \cdot \vec f_i . 
\end{equation}
}

\paragraph*{\textbf{The special case of momentum affinity  (the pressure tensor)}}

The case of momenta $m v_{\alpha}$ affinities needs again to be discussed apart. The flux of momentum is a second order tensor whose components are $p_{\alpha}v_{\beta}$ and it is of common use to express it in terms of the pressure tensor $P_{\alpha\beta}$, which quantifies the change of momentum along the direction $\alpha$ across a unit surface normal to the direction $\beta$. Another second order tensor of interest is the so-called velocity gradient tensor $\partial_{\alpha} v_{\beta}$, which may be decomposed into three tensors:
\begin{equation}
    \label{eq:velocity_tensor}
    \partial_{\beta} v_{\alpha} = \left(\frac{1}{2}\left(  \partial_{\beta} v_{\alpha} + \partial_{\alpha} v_{\beta} \right) - \frac{1}{3} (\nabla \cdot \vec v)  \delta_{\alpha,\beta}\right) + \frac{1}{2}\left(  \partial_{\beta} v_{\alpha} - \partial_{\beta} v_{\alpha} \right) + \frac{1}{3} (\nabla \cdot \vec v) \delta_{\alpha\beta}.
\end{equation}
The first term in the right hand side of Eq.~\eqref{eq:velocity_tensor} is the traceless symmetrical velocity gradient tensor, the second term is the antisymmetrical velocity gradient tensor, while the last term represents the isotropic volume change induced by the flow. 
{For writing more compact equations, in the following we will denote tensors with a bold sans serif font, e.g. $\ten T$; the transpose of $\ten T$ will be denoted $\ten T^+$; the exterior product between $\ten S$ and $\ten T$ will be denoted: $\ten S \ten T$, e.g. $(\ten T \vec v)_{ikl} = T_{ik} v_l$; the interior product will be denoted with a dot, e.g. $(\ten T \cdot \vec v)_{i} = \sum_k T_{ik} v_k$; the scalar product between two tensors of rank 2 will be denoted with a colon, e.g. $\ten S : \ten T = \sum_{i,k} S_{ik} T_{ik}$. The expression of the velocity gradient tensor can then be written:}
\begin{equation}
    \vec \nabla \vec v = \left(\frac{1}{2}\left(\vec \nabla \vec v + \vec \nabla \vec v^{+}\right) - \frac{1}{3}(\vec \nabla \cdot \vec v) \bf{1}  \right)   +   \frac{1}{2}\left(\vec \nabla \vec v - \vec \nabla \vec v^{+}\right) + \frac{1}{3}(\vec \nabla \cdot \vec v) \ten{1} , 
\end{equation}
where we have introduced the unit tensor $\ten{1}$. 

To make the connection between the pressure tensor and the velocity gradient tensor defined in Eq.~\eqref{eq:velocity_tensor}, we may discuss the Onsager coefficient $L_{\vec p\vec p}$ relating the tensor $P_{\alpha\beta}$
to the velocity gradient $\partial_{\delta} v_{\gamma}$.  As it relates two tensors of rank two, it is a tensor of rank $4$ and is characterized therefore by four indices $L_{\vec p\vec p} \equiv L_{\alpha\beta\delta\gamma}$. However, for a bulk \emph{isotropic} fluid, the tensor $L_{\vec p\vec p}$ should be invariant under any arbitrary rotations and symmetry inversion. There are only two tensors which possess this property, $\delta_{\alpha\beta}\delta_{\gamma\delta}+\delta_{\alpha\delta}\delta_{\beta\gamma}$ and $\delta_{\alpha\beta}\delta_{\gamma\delta}$. Therefore, the tensor 
$L_{\vec p\vec p}$ should write~:~$L_{\vec p\vec p\alpha\beta}=
L (\delta_{\alpha\beta}\delta_{\gamma\delta}+\delta_{\alpha\delta}\delta_{\beta\gamma}) + l (\delta_{\alpha\beta}\delta_{\gamma\delta})$. 
Additionally, we note that in an equilibrium situation, the pressure tensor is not vanishing: $P_{\alpha\beta} = P_{\rm eq}\delta_{\alpha\beta}$, 
with $P_{\rm eq}=P_{\rm eq}(T,\rho)$ the equilibrium pressure, which is a function of the system temperature $T$ and density $\rho$ through the equation of state of the fluid. 
In the following, we will therefore consider the \emph{change} of pressure induced by the flow, through the so-called hydrodynamic stress tensor defined as:
\begin{equation}
    \label{eq:definition_stress_tensor}
    - \sigma_{\alpha\beta}=P_{\alpha\beta}-P_{\rm eq}\delta_{\alpha\beta} .
\end{equation}
One can then write, for an isothermal flow, for which 
$\partial_{\alpha}(v_{\beta}/T)=(\partial_{\alpha} v_{\beta})/T$:
\begin{equation}
\label{eq:pressure_tensor_velocity_gradient}
    \ten \sigma = \frac{L}{T}\left(\frac{1}{2}\left(\vec \nabla \vec v + \vec \nabla \vec v^{+}\right) - \frac{1}{3}(\vec \nabla \cdot \vec v) \ten{1}  \right) + \frac{l}{T} \frac{1}{3}(\vec \nabla \cdot \vec v) \ten{1} . 
\end{equation}
where $T$ denotes the temperature and we have introduced two constants $L$ and $l$ whose interpretation will be given below. 

It is also important to note that the hydrodynamic stress tensor $\vec \sigma$ 
is proportionnal to the symmetric velocity gradient tensor, Eq.~\eqref{eq:velocity_tensor}. This is a consequence of the assumed isotropy of the fluid, which implies also that the pressure tensor is symmetric, $P_{\alpha\beta}=P_{\beta\alpha}$. 

Equation~\eqref{eq:pressure_tensor_velocity_gradient} may become more familiar to the reader if we consider some specific examples of hydrodynamic flows. First, imagine a shear flow along the $x$ direction with $z$ the velocity gradient direction, $\vec v = v(z) \vec e_x$. The pressure tensor takes the form:
\begin{equation}
\label{eq:Newton_equation_shear}
\sigma_{xz} 
=\eta \frac{\partial v}{\partial z}, 
\end{equation}
where $\eta=L/2T$ is the fluid shear viscosity.

Compressive velocity fields $\vec v=v(x) \vec e_x$ constitute another important type of flows, for which Eq.~(\ref{eq:pressure_tensor_velocity_gradient}) writes:
\begin{equation}
\label{eq:Newton_equation_bulk}
\sigma_{xx} 
=\zeta \frac{\partial v}{\partial x}, 
    \end{equation}
where $\zeta=l/T$ is the fluid bulk viscosity. This latter quantity plays a role in situations where the flow is compressible. 
In all the following, however, we will mainly deal with flows which may be considered as incompressible and therefore we will neglect the bulk viscosity.

\subsubsection{Onsager matrix in a nanofluidic channel}

In the following, we will precise the form of the Onsager response matrix for a mixture of $N$ species (typically a solvent containing $N-1$ solute species), confined in a nanofluidic channel. To simply the equations, we neglect any chemical reaction.

\paragraph*{\textbf{Conservation equations}}

Before presenting the response matrix, it is useful to write the conservation equations for the system. 
First, one can write the conservation equations obeyed by the species $n$, relating their \emph{number density} $\rho_n$ and their velocity $\vec v_n$:
\begin{equation}
    \label{eq:conservation_rho_n}
    \frac{\partial \rho_n}{\partial t} = -\vec \nabla \cdot \vec j_n, 
\end{equation}
where $\vec j_n=\rho_n \vec v_n$ is the flux of species $n$. 
From Eq.~\eqref{eq:conservation_rho_n}, using $\rho=\sum_n m_n \rho_n$ (with $m_n$ the particle mass of species $n$), and defining the barycentric velocity $\vec v = \sum_n m_n \rho_n \vec v_n /\rho$, one obtains the mass conservation equation:
\begin{equation}
\label{eq:conservation_rho}
    \frac{\partial \rho}{\partial t} = -\vec \nabla \cdot \left( \rho \vec v \right). 
\end{equation}
Introducing the (barycentric) substancial time derivative $\dd/\dd t = (\partial/\partial t + \vec v \cdot \vec \nabla)$ and the specific volume $v=1/\rho$, Eq.~\eqref{eq:conservation_rho} can be recast into: 
\begin{equation}
\label{eq:conservation_rho_2}
    \frac{\dd \rho}{\dd t} = - \rho \vec \nabla \cdot \vec v, \quad \text{or} \quad  \rho \frac{\dd v}{\dd t} = \vec \nabla \cdot \vec v . 
\end{equation}
As a side note, using the mass conservation equation, one can write a useful relation for any local property $a$: 
\begin{equation}
\label{eq:relation_dd_partial}
\frac{\partial (\rho a)}{\partial t} = \rho \frac{\dd a}{\dd t} - \vec \nabla \cdot (\rho a \vec v) ,
\end{equation}
where $\rho a \vec v$ is the convective flux of $a$.

From Eq.~\eqref{eq:conservation_rho_n} and Eq.~\eqref{eq:conservation_rho}, one can write another conservation equation relating the mass fractions $c_n=m_n \rho_n/\rho$ and the diffusion fluxes $\delta \vec j_n=\rho_n (\vec v_n - \vec v)$, i.e. the fluxes of species $n$ in the barycentric frame: 
\begin{equation}
\label{eq:conservation_c_n}
\rho \frac{\dd c_n}{\dd t} = 
- \vec \nabla \cdot \left( m_n \delta \vec j_n \right),   
\end{equation}
Note that by construction, $\sum_n m_n \delta \vec j_n = \vec 0$, so that only $N-1$ of the $N$ equations, Eq.~\eqref{eq:conservation_c_n}, are independent. 
From Eq.~\eqref{eq:conservation_rho_n}, one can also derive the charge conservation equation, relating the charge density $\rho_{\rm e}=\sum_n q_n \rho_n$ ($q_n$ being the charge carried by the species $n$) and the electric current $\vec j_{\rm e}=\sum_n q_n \rho_n \vec v_n = \sum_n q_n \vec j_n$:
\begin{equation}
    \label{eq:conservation_rho_e}
    \frac{\partial \rho_{\rm e}}{\partial t} = -\vec \nabla \cdot \vec j_{\rm e}. 
\end{equation}

Momentum conservation writes:
\begin{equation}
\label{eq:conservation_momentum}
\rho \frac{\dd \vec v}{\dd t} = 
- \vec \nabla \cdot \ten P + \rho_{\rm e} \vec E, 
\end{equation}
where we have introduced the external electric field $\vec E$, and $\ten P = P_{\rm eq} \ten 1 - \ten \sigma$ is the previously defined pressure tensor. 
Using Eq.~\eqref{eq:relation_dd_partial}, momentum conservation can also be written: 
\begin{equation}
\label{eq:conservation_momentum_partial}
\frac{\partial (\rho \vec v)}{\partial t} = - \vec \nabla \cdot (\rho \vec v \vec v + \ten P) + \rho_{\rm e} \vec E. 
\end{equation}
From Eq.~\eqref{eq:conservation_momentum}, one can derive a balance equation for the kinetic energy of the center of mass: 
\begin{equation}
\label{eq:conservation_KEcom}
\frac{\partial \left( \frac{1}{2}\rho \vec v^2\right)}{\partial t} = 
-\vec \nabla \cdot \left( \frac{1}{2}\rho\vec v^2 \vec v + \ten P \cdot \vec v \right) 
+ \ten P : \vec \nabla \vec v + \rhoe \vec v \cdot \vec E ,  
\end{equation}
with $\ten P : \vec \nabla \vec v = P_{\rm eq} \vec \nabla \cdot \vec v - \ten \sigma : \vec \nabla \vec v$. 
If the electric field derives from an electric potential $V$ independent of time, $\partial V / \partial t = 0$, one can write a balance equation for the electrostatic energy: 
\begin{equation}
\label{eq:conservation_ES}
\frac{\partial \left( \rhoe V \right)}{\partial t} = 
-\vec \nabla \cdot \left( V \vec j_\text{e} \right) - \vec j_\text{e} \cdot \vec E . 
\end{equation}

Finally, energy \emph{conservation} takes the form:
\begin{equation}
\label{eq:conservation_energy}
\frac{\partial \left(\rho e \right)}{\partial t} = - \vec \nabla \cdot  \vec j_{\rm en}, 
\end{equation}
with $e$ is the specific energy of the system (i.e. the energy per unit mass), and $\vec j_{\rm en}$ the total energy flux.
The energy can be written: 
\begin{equation}
     \label{eq:energy_density}
    \rho e = \frac{1}{2} \rho \vec v^2 + \rho u + \rhoe V;
\end{equation}
the first term on the right hand side is kinetic energy of the center of mass, $u$ is the specific internal energy and the last term is the electrostatic energy. 
The energy flux writes: 
\begin{equation}
\label{eq:energy_flux}
    \vec j_{\rm en} = \rho e \vec v + \ten P \cdot \vec v 
    + \delta \vec j_{\rm e} V + \delta \vec j_q,   
\end{equation}
where $\delta \vec j_{\rm e} = \vec j_{\rm e} - \rhoe \vec v = \sum_n q_n \delta \vec j_n$ is the diffusion electric current. 
The total energy flux includes a convective term, a contribution due to the mechanical work performed on the system, and an electric energy flux {due to the diffusion of charged particles in the electric potential field}; by definition, the remaining term $\delta \vec j_q$ is the (non-convective) heat flux.

Combining Eqs.~\eqref{eq:conservation_KEcom} to \eqref{eq:energy_flux},
one can express the conservation of the specific internal energy $u$ as:
\begin{equation}
   \label{eq:conservation_internal_energy}
\frac{\partial \left( \rho u \right)}{\partial t} = -\vec \nabla \cdot \left(\rho u \vec v + \delta \vec j_q \right) - \ten P : \vec \nabla \vec v + \delta \vec j_{\rm e} \cdot \vec E.   
\end{equation}
{The internal energy flux in the divergence includes a convective term, $\rho u \vec v$, and the (non-convective) heat flux $\delta \vec j_q$; the right hand side of Eq.~\eqref{eq:conservation_internal_energy} also includes production of internal energy due to viscous dissipation and to Joule effect.}

\paragraph*{\textbf{Entropy production and Onsager matrix}}

Now, we are in a position to compute the local entropy production. 
Denoting $s$ the specific entropy (i.e. the entropy per unit mass), $\vec j_s$ the entropy flux (per unit surface and unit time), and $\sigma_s$ the entropy production (per unit volume and unit time), the second principle of thermodynamics can be written locally: 
\begin{equation}
    \label{eq:second_principle}
    \frac{\partial (\rho s)}{\partial t}=-\vec \nabla \cdot \vec j_s + \sigma_s,   
\end{equation}
with $\sigma_s \ge 0$. 
This equation can be rewritten: 
\begin{equation}
    \label{eq:second_principle_2}
    \rho \frac{\dd s}{\dd t}=-\vec \nabla \cdot \delta \vec j_s + \sigma_s, 
\end{equation}
where $\delta \vec j_s = \vec j_s - \rho s \vec v$ is the non-convective entropy flux. 

Under the local thermal equilibrium assumption, one can apply the thermodynamic identity: 
\begin{equation}
    \label{eq:LTE}
    T \dd s = \dd u + P_{\rm eq} \dd v - \sum_n \frac{\mu_n}{m_n} \dd c_n, 
\end{equation}
with $v=1/\rho$ the specific volume. 
In particular, we assume that Eq.~\eqref{eq:LTE} remains valid for a mass element followed along its motion: 
\begin{equation}
    \label{eq:LTE_2}
    T \frac{\dd s}{\dd t} = \frac{\dd u}{\dd t} + P_{\rm eq} \frac{\dd v}{\dd t} - \sum_n \frac{\mu_n}{m_n} \frac{\dd c_n}{\dd t}.  
\end{equation}
Finally, combining the conservation equations presented above, one can derive expressions for $\delta \vec j_s$ and $\sigma_s$ in Eq.~\eqref{eq:second_principle_2}: 
\begin{equation}
\label{eq:entropy_flux}
\delta \vec j_s = \frac{\delta \vec j_q - \sum_n \mu_n \delta \vec j_n}{T} , 
\end{equation}
and 
\begin{equation}
    \label{eq:entropy_production}
    \sigma_s = 
    - \frac{1}{T} \delta \vec j_q \cdot \frac{\vec \nabla T}{T} 
    + \frac{1}{T} \ten \sigma : \vec \nabla \vec v 
    - \sum_n  \delta \vec j_{n} \cdot \vec \nabla \left(\frac{\mu_n}{T}\right)
    - \frac{1}{T} \delta \vec j_{\rm e} \cdot \vec \nabla V .     
\end{equation}

\subparagraph*{\textbf{Stationary flows}}
For a laminar, stationary 
flow, one can rewrite the hydrodynamic contribution to the entropy production. Indeed, the conservation of momentum gives $-\vec \nabla \cdot \ten \sigma = -\vec \nabla P_{\rm eq} +\rhoe \vec E$, so that: 
\begin{equation}
    \label{eq:essai2a}
    \frac{1}{T} \ten \sigma : \vec \nabla \vec v = 
    \vec \nabla \cdot \left( \frac{\ten \sigma \cdot \vec v}{T} \right)
    + \left( \frac{\ten \sigma \cdot \vec v}{T} \right) \cdot \frac{\vec \nabla T}{T}
    - \frac{\vec v}{T} \cdot \vec \nabla P_{\rm eq} + \frac{\vec v}{T} \cdot \rhoe \vec E.
\end{equation}
One can then rewrite the entropy balance equation with a new definition of the entropy flux: 
\begin{equation}
\label{eq:entropy_flux_flow}
\delta \vec j_s^\text{flow} = \frac{\delta \vec j_q - \sum_n \mu_n \delta \vec j_n - \ten \sigma \cdot \vec v}{T} , 
\end{equation}
where the entropy production now writes (using $\vec j_e = \delta \vec j_e + \rhoe \vec v$): 
\begin{equation}
    \label{eq:entropy_production_flow}
    \sigma_s = 
    - \frac{1}{T} \left(\delta \vec j_q - \ten \sigma \cdot \vec v \right) \cdot \frac{\vec \nabla T}{T} 
    - \sum_n  \delta \vec j_{n} \cdot \vec \nabla \left(\frac{\mu_n}{T}\right)
    - \frac{1}{T}\, \vec v \cdot \vec \nabla P_{\rm eq}
    - \frac{1}{T}\, \vec j_{\rm e} \cdot \vec \nabla V .      
\end{equation}
To simplify this expression further, 
using $h=u+P_{\rm eq}/\rho$, one can write: 
\begin{equation}
    \label{eq:enthalpy_variation_stationary_flow}
    \rho \frac{\dd h}{\dd t}=\rho \frac{\dd u}{\dd t}+\vec v \cdot \vec \nabla P_{\rm eq} + P_{\rm eq} \vec \nabla \cdot \vec v . 
\end{equation}
Using Eq.~\eqref{eq:conservation_internal_energy}, and $-\vec \nabla \cdot \ten \sigma = -\vec \nabla P_{\rm eq} +\rhoe \vec E$, one obtains: 
\begin{equation}\label{eq:enthalpy_flux_1} 
    \rho \frac{ \dd h}{\dd t}=-\vec \nabla \cdot (\vec \delta \vec j_q - \ten \sigma \cdot \vec v) + \vec j_e \cdot \vec E, 
\end{equation}
so that $\vec \delta \vec j_q - \ten \sigma \cdot \vec v$ can be identified with the non-convective enthalpy flux $\delta \vec j_h = \sum_n h_n m_n \delta \vec j_n$, with $h_n$ the partial specific enthalpies, and 
the entropy production simplifies into: 
\begin{equation}
    \label{eq:entropy_production_flow_2}
    \sigma_s = 
    - \frac{1}{T}\, \vec v \cdot \vec \nabla P_{\rm eq}
    - \frac{1}{T}\, \vec j_{\rm e} \cdot \vec \nabla V       
    - \sum_{n=1}^{N}  \delta \vec j_{n} \cdot \vec \nabla \left(\frac{\mu_n}{T}\right)
    - \frac{1}{T} \delta \vec j_h \cdot \frac{\vec \nabla T}{T} .
\end{equation}
As a side note, 
defining the partial specific entropies $s_n=-(\mu_n/m_n-h_n)/T$, the entropy flux also simplifies into: $\delta \vec j_s^\text{flow} = (\delta \vec j_h - \sum_n \mu_n \delta \vec j_n)/T  = \sum_n s_n m_n \delta \vec j_n$. 
The expression of the entropy production, Eq.~\eqref{eq:entropy_production_flow_2}, is of the form Eq.~\eqref{eq:form_bilinear_identification_fluxes}, which can be used to establish the flux-force relations when studying coupled effects involving hydrodynamic, electrical, chemical and thermal transport: 
\begin{equation}
 \begin{pmatrix}
 \vec v\\
 \vec j_{\rm e}\\
 \delta \vec j_{1}\\
 \vdots \\
 \delta \vec j_{N}\\
 \delta \vec j_h
 \end{pmatrix}
 = 
 \begin{pmatrix}
 L_{qq}/T & L_{qe}/T & L_{q1} & \cdots & L_{qN} & L_{qh}/T \\
 L_{eq}/T & L_{ee}/T & L_{e1} & \cdots & L_{eN} & L_{eh}/T \\
 L_{1q}/T & L_{1e}/T & L_{11} & \cdots & L_{1N} & L_{1h}/T \\
 \vdots & \vdots & \vdots & \ddots & \vdots & \vdots \\
 L_{Nq}/T & L_{Ne}/T & L_{N1} & \cdots & L_{NN} & L_{Nh}/T \\
 L_{hq}/T & L_{he}/T & L_{h1} & \cdots & L_{hN} & L_{hh}/T 
 \end{pmatrix}
 \begin{pmatrix}
 -\vec \nabla P_{\rm eq}\\
 -\vec \nabla V\\
 -\vec \nabla (\mu_1/T) \\
 \vdots \\
 -\vec \nabla (\mu_N/T) \\
 -\vec \nabla T / T 
 \end{pmatrix} . 
\label{eq:Onsager_responsematrix}
\end{equation}

A different form of the entropy production can be obtained by using $T \vec \nabla (\mu_n/T) = \vec \nabla \mu_n - (\mu_n/T) \vec \nabla T$, and the definition of the entropy flux, Eq.~\eqref{eq:entropy_flux_flow}: 
\begin{equation}
    \label{eq:entropy_production_flow_3}
    \sigma_s = 
    - \frac{1}{T}\, \vec v \cdot \vec \nabla P_{\rm eq}
    - \frac{1}{T}\, \vec j_{\rm e} \cdot \vec \nabla V       
    - \frac{1}{T}\, \sum_{n=1}^{N}  \delta \vec j_{n} \cdot \vec \nabla \mu_n
    - \frac{1}{T}\, \delta \vec j_s^\text{flow} \cdot \vec \nabla T .
\end{equation}
With this choice of flux-force couples, the Onsager matrix writes: 
\begin{equation}
 \begin{pmatrix}
 \vec v\\
 \vec j_{\rm e}\\
 \delta \vec j_{1}\\
 \vdots \\
 \delta \vec j_{N}\\
 \delta \vec j_s^\text{flow}
 \end{pmatrix}
 = 
 \begin{pmatrix}
 L_{qq}'/T & L_{qe}'/T & L_{q1}'/T & \cdots & L_{qN}'/T & L_{qh}'/T \\
 L_{eq}'/T & L_{ee}'/T & L_{e1}'/T & \cdots & L_{eN}'/T & L_{eh}'/T \\
 L_{1q}'/T & L_{1e}'/T & L_{11}'/T & \cdots & L_{1N}'/T & L_{1h}'/T \\
 \vdots & \vdots & \vdots & \ddots & \vdots & \vdots \\
 L_{Nq}'/T & L_{Ne}'/T & L_{N1}'/T & \cdots & L_{NN}'/T & L_{Nh}'/T \\
 L_{hq}'/T & L_{he}'/T & L_{h1}'/T & \cdots & L_{hN}'/T & L_{hh}'/T 
 \end{pmatrix}
 \begin{pmatrix}
 -\vec \nabla P_{\rm eq}\\
 -\vec \nabla V\\
 -\vec \nabla \mu_1 \\
 \vdots \\
 -\vec \nabla \mu_N \\
 -\vec \nabla T 
 \end{pmatrix} . 
\label{eq:Onsager_responsematrix_2}
\end{equation}
{Note however that the entropy flux $\delta \vec j_s^\text{flow}$ quantifying now the thermal response is difficult to evaluate, so that this form of the Onsager matrix is less suitable to study coupled effects involving thermal transport.}

{In these equations, one can eliminate the gradient of chemical potential of one constituent -- typically the solvent -- using the Gibbs-Duhem relation, 
$\dd P_{\rm eq} = \rho s \dd T + \sum_n \rho_n \dd \mu_n$, 
which, for a stationary flow and considering gradients along the flow direction, rewrites: 
\begin{equation}
    \vec \nabla \mu_1 = \frac{1}{\rho_1} \vec \nabla P_{\rm eq} - \frac{\rho s}{\rho_1} \vec \nabla T - \sum_{n = 2}^{N} \frac{\rho_n}{\rho_1} \vec \nabla \mu_n .
\end{equation}
It is also possible to group all the solute terms, e.g. defining $\vec \nabla \mu = \sum_{n\ge 2} \vec \nabla \mu_n$ and redefining $\delta \vec j_n = \sum_{n'\ge 2} \delta \vec j_{n'}$. 
}

\paragraph*{\textbf{Discussion}}

{Finally, we would like to discuss the consequences of symmetries on the transport coefficients in 
two important situations. First, let us consider a \emph{bulk} isotropic system. By virtue of Curie principle \cite{Curie1908}, ``the elements of symmetry of the causes should be retrieved in the effects they give rise to''. One of the consequences of Curie's principle is that, for a bulk isotropic system, the gradient of a scalar quantity can only generate the flux of a scalar quantity, and the gradient of a vectorial quantity can only generate the flux of a vectorial quantity. In particular, this implies that, in bulk, a flow can not be generated by a gradient of temperature, chemical potential or electrostatic potential, i.e. $L_{qe} = L_{qn} = L_{qh} = 0$. 
By symmetry, a gradient of the pressure tensor can not generate an electric current, an excess flux of solutes or a heat flux, which implies that $L_{eq} = L_{nq} = L_{hq} = 0$. 
} 

{For a confined system, however, the situation is different, as confinement breaks the isotropy. 
Imagine for instance a nanoslit having a square section: this system has symmetry along the directions parallel to the channel slits, and no symmetry in the direction perpendicular. In this geometry, a longitudinal gradient of temperature may generate a flow along the channel~:~this is the phenomenon we know as \emph{thermo-osmosis}. Similar phenomena include \emph{diffusio-osmosis} and \emph{electro-osmosis}, which describe the flow generated by a longitudinal gradient of solute or by a longitudinal electric field, see Section~\ref{sec:osmotic_flows}. Importantly here, these osmotic phenomena arise because the system has symmetry axis along the channel slit. In the direction perpendicular, there is no symmetry and for instance a gradient of temperature along this direction does not create a flow.}

\subsection{Origin of coupled transport: specific interactions at interfaces}

Although the framework of linear irreversible thermodynamics provides us with useful general relations between the response coefficients of the nanofluidic system, it does not tell us anything about the mechanisms underlying electrokinetic effects. 
As discussed in the previous section, interfaces are key to coupled phenomena such as osmotic flows. In practice, the  
coupling between different types of transport that arises at interfaces is due to the interactions of the fluid with the walls. In this section we will provide a brief overview of the different types of interactions at play for water-solid interfaces, discussing in particular their characteristic scales.  

\subsubsection{Molecular interactions}
\label{subsubsec:molec_interactions}

When working in confined systems, it can be critical to consider the short-range interactions between the water molecules or the solute and the wall, which extend over a few molecular layers. Indeed, in the case of water, the molecular water-wall interactions induce the layering of the liquid at the interface, implying a local change of density close to the wall, and affecting the local viscosity \cite{Botan2011,Hoang2012a}. Also, liquid-solid interactions affect the local atomic internal energy and pressure, hence the local enthalpy \cite{proesmans2019}. Similarly, the interaction energy between the solute and the wall can produce a solute excess or a solute depletion.

\subsubsection{Aqueous electrolytes: the electrical double layer}

When a dielectric is plunged into water, several mechanisms can generate a surface charge, together with an opposite charge carried by ions in the liquid (typical mechanisms include dissociation of surface groups and specific adsorption of charged species) \cite{LyklemaBook,Hunter2001,IsraelachviliBook}. These phenomena are specific to polar liquids such as water,  
for which the Bjerrum length $\lB$ (representing the distance at which the thermal energy is comparable to the electrostatic interaction energy, $\lB \sim 7~$\AA{} for water at $25 ^\circ$C) is comparable to (or smaller than) the interatomic distance \cite{IsraelachviliBook}. 

How water molecules and dissolved ions in a solution electrostatically interact with the surface has been extensively investigated, theoretically from the 19th century \cite{helmholtz1853} and more extensively by Gouy and Chapman at the beginning of the 20th century \cite{Gouy1910,Chapman1913}, and experimentally in the 1980s \cite{Derjaguin1980,Derjaguin1987}. 
Educational presentations of the Gouy-Chapman theory can be found in books \cite{LyklemaBook,Hunter2001,IsraelachviliBook}, book chapters \cite{Andelman1995,Markovich2016a}, and articles \cite{Delgado2007}, discussing in particular applications to nanofluidics \cite{Schoch2008,Bocquet2010,Hartkamp2018,Kavokine2021}; 
useful equations for the description of slit and cylindrical channels are gathered in an online formulary \cite{Herrero2021}. 
We recall simply here the main ingredients of this theory, and  discuss its limitations.

\begin{figure}
\begin{subfigure}{0.47\linewidth}
    \centering
    \includegraphics[width=0.99\linewidth]{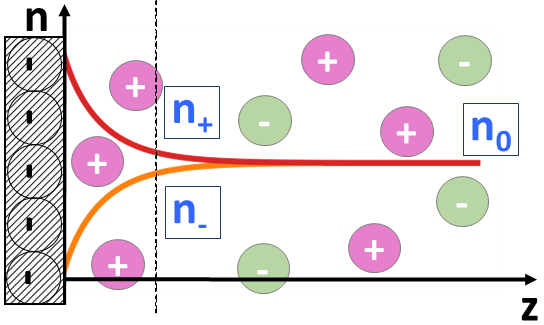}
    \caption{}
    \label{fig:PBframework_ndens}
\end{subfigure}
\begin{subfigure}{0.49\linewidth}
    \centering
    \includegraphics[width=0.99\linewidth]{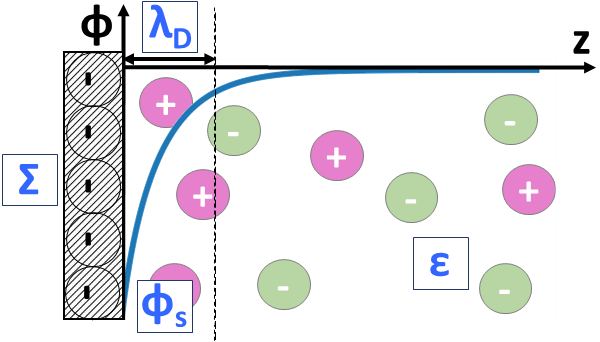}
    \caption{}
    \label{fig:PBframework_phi}
\end{subfigure}
\caption{Local picture of the electrical double layer at a charged wall: (a) profiles of cation $n_+$ and anion $n_-$ densities, with $n_0$ their bulk value; (b) profile of reduced electric potential $\phi$, with $\phis$ its value at the wall.}
\label{fig:PB_framework}
\end{figure}

Let us consider a charged surface in contact with an aqueous electrolyte. Far from the interface, the positive and negative ions are dispersed in the solution, due to entropy, and will have the same concentration. Nevertheless, close to the wall, ions with a charge of opposite sign to that of the surface (counter-ions) will accumulate, and ions with a charge of the same sign as the surface (co-ions) will be depleted, forming the so-called electrical double layer (EDL), see Fig.~\ref{fig:PB_framework}. 

Gouy and Chapman described the ion distribution near the wall, and the subsequent electrostatic potential induced by the surface charge, by combining the Poisson equation for electrostatics and the Boltzmann distribution of the ions into the so-called Poisson-Boltzmann (PB) equation, under certain assumptions \cite{Andelman1995,Markovich2016a}: 
\begin{itemize}
    \item the Poisson equation is written assuming that the solvent has a local, homogeneous and isotropic dielectric permittivity; 
    \item the Boltzmann distribution of the ions is written assuming that the energy of the ions results only from their Coulomb interactions with the other ions and the wall, described at a mean-field level. 
\end{itemize}

For simplicity, let's consider a smooth charged surface, located at $z=0$, with uniform surface charge density $\Sigma$ (Fig.~\ref{fig:PB_framework}). In this case, the electric potential along the channel $V(z)$ is given by the Poisson equation:
\begin{equation}
    \eps \frac{\dd^2 V}{\dd z^2} = -\rhoe(z),
    \label{eq:Poisson}
\end{equation}
where $\eps$ 
is the solvent dielectric permittivity, 
and $\rhoe$ is the charge density in the liquid. Such charge density can be expressed in terms of the positive $n_+$ and negative $n_-$ ion densities as $\rhoe(z) = q_{e_+} n_+(z) - q_{e_-} n_-(z)$, where $q_{e_\pm} = Z_\pm\, e$ is the absolute ionic charge, with $e$ the elementary charge and $Z_\pm$ the ion valence. 
The ion concentrations in the liquid are given by the Boltzmann equation:
\begin{equation}
    n_\pm(z) = n_0 \, \exp[\mp \beta q_{e_\pm} V(z)],
    \label{eq:Boltzmann}
\end{equation}
where $n_0$ is the ion concentration far from the wall (see Fig.~\ref{fig:PBframework_ndens}) and $\beta=1/(k_\mathrm{B}T)$, with $T$ the temperature and $k_\mathrm{B}$ the Boltzmann factor. Taking into account Eq.~\eqref{eq:Boltzmann}, and if we consider that cations and anions have the same valency, $q_{e_+}=q_{e_-}=q_e$, the charge density can be expressed as $\rhoe(z) = q_e [n_+(z) - n_-(z)] = -2 q_e n_0 \sinh[\beta q_e V(z)]$. Finally, one can substitute this expression obtained from the Boltzmann equation for $\rhoe$ in Eq.~\eqref{eq:Poisson}, resulting in:
\begin{equation}
    \ddiff{\phi(z)}{z} = \frac{2\beta q_e^2 n_0}{\eps} \sinh[\phi(z)] = 8 \pi n_0 \lB \sinh[\phi(z)],
    \label{eq:raw_PB}
\end{equation}
where $\phi(z) = \beta q_e V(z)$ is the reduced potential (see Fig.~\ref{fig:PBframework_phi}). This expression introduces a system characteristic length, the Bjerrum length $\lB = \beta q_e^2/(4 \pi \eps)$, which corresponds to the distance at which the thermal energy is comparable to the electrostatic interaction energy between two ions.

Equation~\eqref{eq:raw_PB} can be rewritten to obtain  the so-called Poisson-Boltzmann equation,
\begin{equation}
    \ddiff{\phi}{z} = \frac{1}{\debye^2} \sinh[\phi(z)],
    \label{eq:PB}
\end{equation}
where a typical screening length, the so-called Debye length has been introduced: 
\begin{equation}
\debye=1/\sqrt{8 \pi \lB n_0}.
\end{equation}
In the case of a planar  smooth wall in contact with an infinite reservoir, assuming a vanishing potential far from the wall, Eq.~\eqref{eq:PB} can be integrated as:
\begin{equation}
    \phi(z) = 4 \text{atanh}\left(\gamma e^{-z/\debye}\right),
    \label{eq:PB_planarwall}
\end{equation}
with $\gamma = \tanh(\phis/4)$, where $\phis$ is the potential at the wall surface. 
Generally, the surface is characterized by its surface charge density $\Sigma$. A relationship between $\phis$ and $\Sigma$ can be  established, considering that at the wall surface, the electric field $E_\text{s}$ reads $E_\text{s} = - \eval{\diff{V}{z}}_{z=0} = \frac{\Sigma}{\eps}$. Then, for a planar  wall, one obtains:
\begin{equation}
    \phis = 2 \, \sgn(\Sigma) \text{asinh}\left( \frac{\debye}{\lGC} \right),
    \label{eq:phi0}
\end{equation}
where $\lGC = q_e /( 2 \pi \lB |\Sigma|)$ is the Gouy-Chapman length. This relation is an alternative form of the so-called \emph{Grahame equation}. From the expression of $\phis$, one can write for $\gamma$:
\begin{equation}
    \gamma = \frac{\sgn(\Sigma)}{\debye/\lGC}\qty( -1 + \sqrt{1+(\debye/\lGC)^2} ).
    \label{eq:gamma_PB}
\end{equation}

When the reduced potential $|\phi|$ is much lower than 1 everywhere, and therefore when $|\phis| = \max(|\phi|) \ll 1$, the PB equation can be linearized; this is the Debye-H\"uckel (DH) regime. In that regime the potential reduces to: 
\begin{equation}
   \phi^\mathrm{DH}(z) = \phis \exp(-\frac{z}{\debye}).
   \label{eq:PB_DH}
\end{equation}
Using Grahame equation, Eq.~\eqref{eq:phi0}, it appears that the DH regime is found when $\debye \ll \lGC$, which occurs for low $\Sigma$ and/or large $n_0$.
The other limit of high surface charge and/or potential, known as the Gouy-Chapman limit (GC), is obtained when $\debye \gg \lGC$, as  detailed in \cite{Andelman1995}.

\paragraph*{\textbf{Validity of the hypotheses and limitations}}

We can now discuss the validity of this modeling and in particular the limits of the different hypotheses underlying the Gouy-Chapman theory. Let's first consider the mean-field approximation. Ionic correlations can  be discarded if the typical Coulombic interaction energy between two ions is small compared to $\kt$, which reads if we introduce the so-called plasma parameter $\Gamma$ \cite{Levin2002,Levin2003,Joly2006},
\begin{equation}
    \Gamma = \frac{\beta q^2}{4 \pi \eps d_\mathrm{ions}} = \frac{\lB}{d_\mathrm{ions}} < 1,
    \label{eq:plasma_gal}
\end{equation}
where $d_\mathrm{ions}$ is the typical inter-ionic distance.
At the surface, $1/d_\mathrm{ions}^2=|\Sigma|/q_e$, and we can rewrite $\Gamma = \sqrt{|\Sigma| \lB^2 / q_e}$.
From this, there is a critical charge  density above which ionic correlations must be considered,
\begin{equation}
    \abs{\Sigma^\mathrm{c}} = \frac{q_e}{\lB^2} = \frac{(4\pi\eps/\beta)^2}{(Ze)^3}, 
\end{equation} 
where the last expression highlights the strong impact of ion valence on $\abs{\Sigma^\mathrm{c}}$.
For a monovalent salt in water at $300\,$K, $\abs{\Sigma^\mathrm{c}} \sim 330\,$mC/m$^2$, but for divalent ions it drops to $\sim 40\,$mC/m$^2$. 

Similarly, a critical concentration can be determined in the bulk; with $d_\mathrm{ions} = \left(2 n_0^\mathrm{c} \right)^{-1/3}$, one obtains
\begin{equation}
    n_0^\mathrm{c} = \frac{1}{2\, \lB^3} = \frac{(4\pi\eps/\beta)^3}{2(Ze)^6}. 
\end{equation}
For a monovalent salt in water at $300\,$K, $n_0^\mathrm{c}\sim 2\,$M, but for a divalent salt it drops to $\sim 30$\,mM. 
Finally, it is important to note that, for monovalent ions in water at room temperature, $\lB \sim 7\,$\AA{} is greater than the ionic size, so that there will be no steric repulsion effects when $\Gamma = \lB/d_\mathrm{ions} < 1$. 

Other limits of the Gouy-Chapman theory, related to the molecular detail of the liquid-solid interface, will be discussed in Sec.~\ref{ssec:beyond_standard}.
Nevertheless, the validity of the presented model has been assessed by numerous experiments and simulations under standard conditions and typically for symmetric monovalent salts dissolved in water, which are systems that can be easily found in everyday media (for example sea water), and with a deep theoretical interest due to its multiple and promising applications, as the ones discussed in the following.

\section{Osmotic flows}
\label{sec:osmotic_flows}

\begin{figure}
    \centering
    \includegraphics[width=0.8\linewidth]{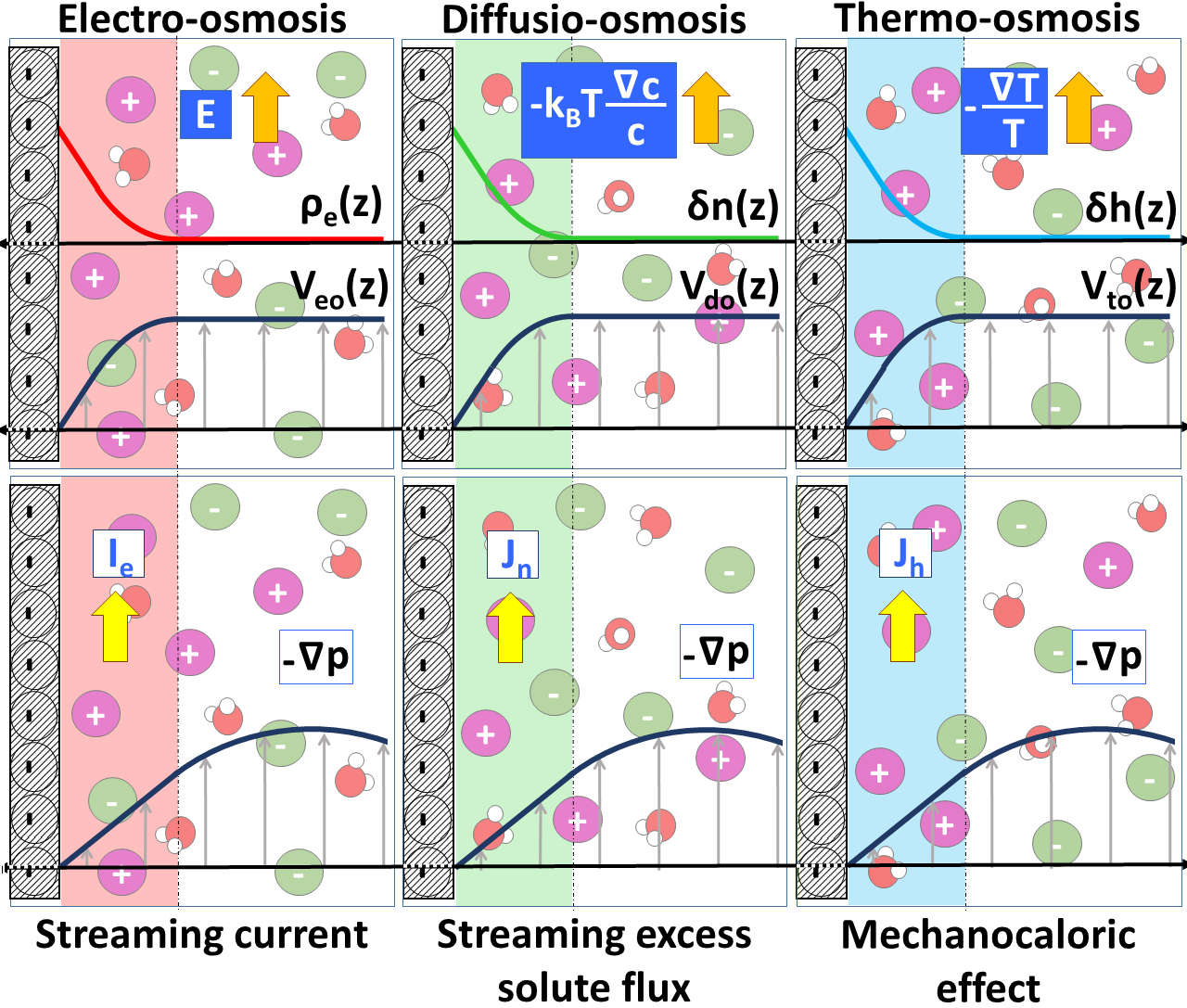}
    \caption{Schematics of the different osmotic flows considered in this chapter, and their Onsager reciprocal effects; Left: electro-osmosis (top), the flow induced by an electric field, and streaming current (bottom), the electric current generated by a pressure gradient; Middle: diffusio-osmosis (top), the flow induced by a solute concentration gradient, and streaming excess solute flux (bottom), the solute flux generated by a pressure gradient, to which the flux expected from advection of the bulk solute concentration by the pressure driven flow is removed; Right: thermo-osmosis (top), the flow induced by a temperature gradient, and mechanocaloric effect (bottom), the heat flux generated by a pressure gradient. The reciprocal effects are quantified by the same response coefficient, which is controlled by the interfacial charge excess, solute excess, and enthalpy excess for electro-, diffusio-, and thermo-osmosis respectively.}
    \label{fig:osmosis}
\end{figure}

From linear irreversible thermodynamics (see section~\ref{sec:irreversible_thermodynamics}), it is expected that flows can be generated at surfaces by non-hydrodynamical thermodynamic gradients applied along the interface. Such flows are called osmotic flows, and are illustrated in Fig.~\ref{fig:osmosis}, together with their reciprocal effects in the Onsager response matrix. Beyond the well-known electro-osmosis (generated by electric fields along the wall), Fig.~\ref{fig:osmosis} also illustrates diffusio-osmosis (generated by solute concentration gradients), thermo-osmosis (generated by thermal gradients), and their reciprocal effects.

In this section we will explore the origin of osmotic flows, 
showing how they are controlled by a thin layer of liquid in the vicinity of the wall where the liquid properties differ from their bulk values due to interactions with the wall, which we will refer to as the \emph{interaction layer}.

\subsection{General description of osmotic flows}

In this section, we will derive a general expression for the osmotic velocity profile close to a planar wall (Sec.~\ref{sec:vosm}), as a function of the force density profile in the interaction layer. To that aim, we will use Stokes equation (Sec.~\ref{sec:stokes}), with appropriate hydrodynamic boundary conditions (Sec.~\ref{sec:navier}).

\subsubsection{Low-Reynolds hydrodynamics: Stokes equation}

\label{sec:stokes}

The flow of a newtonian, incompressible fluid (with density $\rho$ and shear viscosity $\eta$) is described by the Navier-Stokes equations: 
\begin{align}
\label{eq:navier-stokes}
\rho \frac{ \partial \vec{v} }{ \partial t } + \rho \left(\vec{v} \cdot \vec{\nabla}\right)\vec{v} & = - \vec{\nabla}p + \eta \Delta \vec{v} + \vec{f}_\text{ext},
\intertext{with the condition of incompressibility:}
\vec{\nabla}\cdot\vec{v} &= 0,
\end{align}
where $\vec{v}$ is the velocity, $p$ the pressure, and $\vec{f}_\text{ext}$ an external force density. 

In micro and nanofluidic systems, the typical spatial scale $L$ and velocity $U$ are such that the Reynolds number $\mathrm{Re} = \rho UL/\eta$, 
which quantifies the ratio between the inertial term, $\rho \left(\vec{v} \cdot \vec{\nabla}\right)\vec{v} \sim \rho U^2/L$, and the viscous term, $\eta \Delta \vec{v} \sim \eta U/L^2$, is much smaller than 1. 
In that case, and considering steady-state flows where $\frac{ \partial \vec{v} }{ \partial t } = 0$, Eq.~\eqref{eq:navier-stokes} is reduced to the (much simpler) Stokes equation:
\begin{equation}
\label{eq:stokes}
- \eta \Delta \vec{v} = - \vec{\nabla}p + \vec{f}_\text{ext}.
\end{equation}
Flows described by Eq.~\eqref{eq:stokes} are usually referred to as creeping flows. In the following of the chapter, we will only consider such creeping flows. 
Specifically, we will focus on flows induced in the $x$ direction by external actuation along the same axis, of liquids confined by walls perpendicular to the $z$ direction. In that case, denoting $v$ the velocity along $x$, and $\nabla = \partial_x$, Eq.~\eqref{eq:stokes} rewrites: 
\begin{equation}
\label{eq:stokes_scalar}
- \eta \partial_z^2 v = - \nabla p + f_\text{ext}.
\end{equation}

\subsubsection{The hydrodynamic boundary condition}
\label{sec:navier}

\begin{figure}
    \centering
    \includegraphics[width=0.6\linewidth]{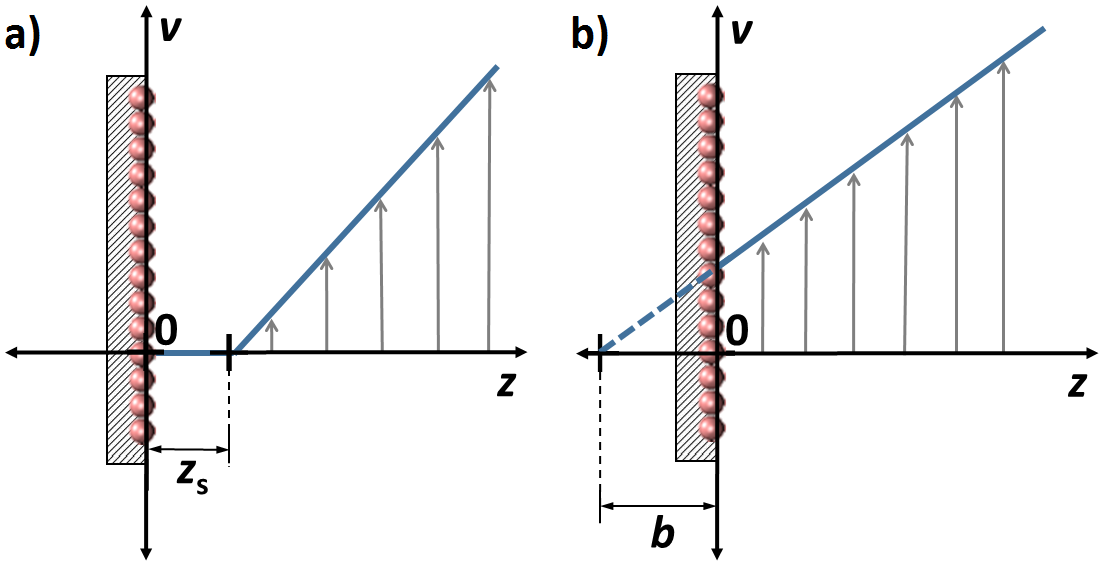}
    \caption{At the nanoscale, the standard no-slip boundary condition (BC) needs to be refined, taking into account either: a) the presence of a stagnant layer with thickness $z_s$; or b) liquid-solid slip with slip length $b$.}
    \label{fig:hbc}
\end{figure}

The standard no-slip boundary condition (BC), which supposes that the fluid velocity vanishes when in contact to the wall, needs to be refined at the nanoscale. Two different situations can occur. 
For strong liquid-solid interactions, the large interfacial friction creates 
a liquid stagnant layer of size $z_\mathrm{s}$ close to the wall (Fig.~\ref{fig:hbc}a), implying a vanishing velocity profile inside the channel. The typical size of the stagnant layer 
is of order of one molecular diameter \cite{Herrero2019}, $\sim 2.75\,$\AA{} for water.

For weak liquid-solid interactions, the small interfacial friction allows for 
an interfacial velocity jump, also known as slip velocity $\vs$, which implies a non vanishing fluid velocity profile at the interface. In this case, as originally done by Navier \cite{Navier1823}, one can suppose that $\vs$ will be proportional to the shear stress $\tau$, which corresponds to the force exerted from the liquid to the solid wall per unit area. Such proportion can be expressed as $\tau = \lambda \vs$, where $\lambda$ is called the friction coefficient. We also know that, supposing a laminar flow far from the wall, $\tau$ will follow Newton's law for viscosity: $\tau = \eta \frac{\partial v_x}{\partial z}$. 
One can match both bulk and interfacial expressions for the shear stress in the so-called ``Navier boundary condition'':
\begin{equation}
    \vs = b \eval{\frac{\partial v_x}{\partial z}}_{z = 0},
    \label{eq:navier}
\end{equation}
where $b= \frac{\eta}{\lambda}$ is the slip length. Although this feature is always present for a confined fluid, it is at the nano- and micro-metric scales that it becomes critical to take it into account, due to the typical order of magnitude of the slip length, $b \sim 10^0-10^2\,$nm  \cite{Bocquet2010,Kavokine2021}.

\subsubsection{Osmotic velocity over a planar wall}
\label{sec:vosm}

In this section, we consider a planar wall, and denote $z$ the distance to the wall, see Fig.~\ref{fig:osmosis}. As detailed in the following for the specific cases of electro-, diffusio-, and thermo-osmosis, a thermodynamic gradient applied along the $x$ direction parallel to the interface will generate a force density $f(z)$ in the interaction layer. We will compute the velocity profile generated by this force density, and the velocity plateau value reached outside the interaction layer, which we will call \textit{osmotic velocity}, and denote $v_\text{osm}^\infty$. 
Integrating Stokes equation, Eq.~\eqref{eq:stokes_scalar}, in the lubrication limit and when no pressure gradient applies, one obtains:
\begin{equation}
    -\eta \frac{\partial v_x}{\partial z}(z)=\int_{+\infty}^z f(z') \,\dd z',
    \label{eq:velocitygrad}
\end{equation}
assuming that the velocity gradient  vanishes far from the wall. Integrating again  between $0$ and  $z$, one obtains:
\begin{equation}
    v_x(z)-v_x(0)= \frac{1}{\eta} \int_0^z \dd z' \int_{z'}^{+\infty}  f(z'') \,\dd z''.
\end{equation}

In the case of slippage, defined by a slip length $b$, one can introduce the relevant boundary condition, Eq.~\eqref{eq:navier}, 
\begin{equation}\label{eq:vosmz_general}
    v_x(z) = \frac{1}{\eta} \left\{ \int_0^{z} \dd z' \int_{z'}^{\infty} f(z'') \,\dd z''  +  b \int_0^{\infty}  f(z) \,\dd z \right\}, 
\end{equation}
which reduces, after an integration by part  and taking its limit far from the wall, to:
\begin{equation}\label{eq:vosm_general}
    v_\text{osm}^\infty = v_x(\infty) = \frac{1}{\eta} \int_0^{\infty} \left( z + b \right) f(z) \,\dd z. 
\end{equation}

Similarly, if we have  a stagnant layer of size $\zs$, Eq.~\eqref{eq:velocitygrad} is integrated between  $\zs$ and $z$, which results in
\begin{equation}\label{eq:vosmz_general_stagnant}
    v_x(z) = \frac{1}{\eta} \int_{\zs}^{z} \dd z' \int_{z'}^{\infty} f(z'') \,\dd z'', 
\end{equation}
which reduces, in the limit far from the wall, in
\begin{equation}\label{eq:vosm_general_stagnant}
    v_\text{osm}^\infty = \frac{1}{\eta} \int_{\zs}^{\infty} \left( z - \zs \right) f(z) \,\dd z. 
\end{equation}

\subsection{Electro-osmosis: flows induced by electric fields}\label{sec:eo}

Electro-osmosis describes a fluid flow generated at a wall under an electrical potential gradient along the wall. This transport mechanism takes its origin in the EDL, where the charge density $\rhoe$ differs from zero. The corresponding response matrix writes: 
\begin{equation}\label{eq:Meo}
 \begin{pmatrix}
 q\\
 j_e
 \end{pmatrix}
 = 
 \begin{pmatrix}
 \cdot & M_\text{eo} \\
 M_\text{eo} & \cdot
 \end{pmatrix}
 \begin{pmatrix}
 f_\text{p}=-\nabla p\\
 E=-\nabla V
 \end{pmatrix}
\end{equation}
with $q$ the flow rate density (average fluid velocity), $j_e$ the electrical current density, $f_\text{p} = -\nabla p$ the pressure gradient, and $E=-\nabla V$ the applied electric field. 
In this section, we will see two possible calculations of the electro-osmotic coefficient $M_\text{eo}$ using the reciprocal formula of Onsager that corresponds to two different fluidic transports : electro-osmotic flow, $q = M_\text{eo} E$, and streaming current, $j_e = M_\text{eo} f_\text{p}$.

\subsubsection{Electro-osmotic flow}

When an external electric field $E$ parallel to the interface is applied, the Coulomb force density, $f(z) = \rhoe (z) E$, puts the fluid into motion. This driving force is limited in the EDL and vanishes outside, where $\rhoe=0$. Moreover, one can use the Poisson equation, Eq.~\eqref{eq:Poisson}, that links the charge density $\rhoe$ with the electric potential $V(z)$ induced by the surface charge, and Grahame equation, Eq.~\eqref{eq:phi0}, which relates the surface potential $\Vs$ to the surface charge $\Sigma$, or equivalently to the Gouy-Chapman length $\lGC$.

From the equations in section~\ref{sec:vosm}, one can then get the electro-osmotic velocity profile, 
in the slip case, 
\begin{equation}
  v_\text{eo} (z) = \frac{\eps E}{\eta} \left( V(z) - \Vs - \frac{\Sigma b}{\eps} \right) = -\frac{\eps E \Vs}{\eta} \left( 1 + \frac{b}{\lambda_\text{eff}} - \frac{V(z)}{\Vs} \right), 
  \label{eq:vEO_gal}
\end{equation}
and in the stagnant layer case, 
\begin{equation}
  v_\text{eo} (z) = \frac{\eps E}{\eta} \left[ V(z) - V(\zs) \right], 
  \label{eq:vEO_gal_noslip}
\end{equation}
defining an effective Debye length $\lambda_\text{eff} = \eps \Vs / \Sigma$, which identifies with $\debye$ in the Debye-H\"uckel limit \cite{Herrero2021}. 
Note that one can equivalently express the electro-osmotic velocity using the reduced potential, the Bjerrum length $\lB$ and the Gouy-Chapman length $\lGC$, e.g. in the slip case: 
\begin{equation}
    v_\text{eo} (z) = \frac{q_e E}{4 \pi \lB \eta} \left[ \phi(z) - \phi_s - 2 \, \sgn(\Sigma) \frac{b}{\lGC} \right] .
    \label{eq:vEO_ch}
\end{equation}

The bulk electro-osmotic velocity, i.e. for $z\gg \debye$, is respectively in the slip case and the stagnant layer case: 
\begin{align}
    \label{eq:veobulk_slip}  
    v_\text{eo}^\infty &= -\frac{\eps}{\eta} \left( \Vs + \frac{\Sigma b}{\eps} \right) \times E = -\frac{\eps \Vs}{\eta} \left( 1 + \frac{b}{\lambda_\text{eff}} \right)  \times E,\\
\label{eq:veobulk_noslip}  v_\text{eo}^\infty &= -\frac{\eps V(\zs)}{\eta} \times E .
\end{align}
When the interaction layers are thin as compared to the channel size, the average fluid velocity identifies with $v_\text{eo}^\infty$, so that the response coefficient writes: $M_\text{eo} = v_\text{eo}^\infty / E$. 
Note that the response coefficient is commonly expressed in terms of 
an effective potential, called $\zeta$ potential and 
defined through the Helmoltz-Smolushowski formula, $v_\text{eo}^\infty = (-\eps \zeta / \eta) E$ -- with $\eps$ and $\eta$ the bulk permittivity and viscosity, respectively, so that $M_\text{eo} = -\eps \zeta / \eta$. From Eqs.~\eqref{eq:veobulk_slip} and \eqref{eq:veobulk_noslip}, one can express $\zeta$, in the slip case and for a stagnant layer, respectively: 
\begin{align}
    \label{eq:zeta_b}
    \zeta &= -\frac{\eta v_\text{eo}^\infty}{\eps E} = \Vs + \frac{\Sigma b}{\eps} = \Vs \left( 1 + \frac{b}{\lambda_\text{eff}} \right), \\
    \label{eq:zeta_zs}
    \zeta &= -\frac{\eta v_\text{eo}^\infty}{\eps E} =  V(\zs). 
\end{align}

The $\zeta$ potential is an important quantity to compute the flow driven by a potential difference. Moreover, it can be used to quantify the reciprocal effect which is a current driven by a pressure difference, called the streaming current.

\subsubsection{Reciprocal effect: streaming current}
\label{sec:streaming_current}

According to Onsager reciprocal relations, the expression of the zeta potential 
can also be derived by considering the reciprocal effect, the streaming current, i.e. the electrical current generated by a pressure gradient, 
in the absence of electric field. 
Here we will focus on slipping walls to obtain Eq.~\eqref{eq:zeta_b}, but the same approach can be followed in the stagnant layer situation, leading to Eq.~\eqref{eq:zeta_zs}. 

We consider a slab channel, with two parallel walls perpendicular to the $z$ axis, located at $z=0$ and $z=d$, see Fig.~\ref{fig:osmosis}; this calculation can be generalized to an arbitrary section, see Ref.~\cite{BarbosaDeLima2017}. We apply a pressure gradient $f_\text{p} = -\nabla p$ along the $x$ direction, which generates a Poiseuille velocity profile $v(z)$. 

The Poiseuille flow is described by Stokes equation: $-\eta \ddiff{v}{z} = f_\text{p}$. The channel is symmetric with regard to $z=d/2$, so that $\left.\diff{v}{z}\right|_{z=d/2} = 0$, and we consider a partial slip BC on the bottom wall: $v(0) = b\,\left.\diff{v}{z}\right|_{z=0}$. 
The resulting flow profile writes: 
\begin{equation}\label{eq:poiseuille}
  v(z) = \frac{f_\text{p}}{2\eta} \left\{ d(z+b) - z^2 \right\}. 
\end{equation}

In the EDL where the liquid is charged, the Poiseuille flow creates a local current density $j_\text{e} = \rho_\text{e} v$. The average electric current density through the channel can be written (by symmetry, one can integrate only over the bottom half of the channel): 
\begin{equation}\label{eq:je_pois}
    j_e = \frac{2}{d} \int_0^{d/2} \rhoe(z) v(z) \,\dd z = - \frac{2\eps}{d} \int_0^{d/2} \ddiff{V}{z} v(z) \,\dd z, 
\end{equation}
where the charge density was replaced using Poisson equation, Eq.~\eqref{eq:Poisson}. 

When the EDLs are thin as compared to the channel height, the integrand in Eq.~\eqref{eq:je_pois} will only differ from zero for $z\ll d$. 
In that case, one can linearize the velocity profile in the integral:     
\begin{equation}\label{eq:vPois_lin}
    v(z)  \simeq \frac{f_\text{p} d}{2 \eta} \left( z + b \right),  
\end{equation}
and one can extend the upper boundary of the integral to infinity: 
\begin{equation}\label{eq:je_pois_lin}
    j_e = - \frac{\eps f_\text{p}}{\eta} \int_0^{\infty} (z+b) \ddiff{V}{z} \,\dd z 
    = - \frac{\eps f_\text{p}}{\eta} \left( \Vs + \frac{\Sigma b}{\eps} \right). 
\end{equation}
Writing that $j_e = (-\eps \zeta / \eta) f_\text{p} \Rightarrow \zeta = -\eta j_e /(\eps f_\text{p})$, one recovers Eq.~\eqref{eq:zeta_b} obtained from the electro-osmotic response.

\subsubsection{Toward giant zeta potentials?} 

Interestingly, in the presence of slip, the $\zeta$ potential is larger than the surface potential.  Therefore, in principle, giant zeta potentials could be obtained by combining a high surface charge and a large slip length. 

In practice, however, large slip lengths appear on hydrophobic surfaces, which are in general only weakly charged; reciprocally, highly charged surfaces are generally hydrophilic and do not slip. The relation between surface charge and slip has been studied by several groups~\cite{Joly2006,Huang2008,Botan2013,Joly2014}. Recently, it has been shown that the charge-slip coupling depends strongly on the charge distribution on the surface \cite{Xie2020}. In particular, it is predicted that giant zeta potentials of $\sim 2000$\,mV could be obtained on polarized graphene, arising from a giant slip length on uncharged graphene, and a favorable charge-slip relation for this extremely smooth surface. One could also think of using superhydrophobic surfaces, which generate giant slip using the fakir effect where the liquid rests on top of nanotextured surfaces. Yet electro-osmosis is not amplified on such surfaces: electro-osmosis is generated locally, and the liquid-vapor interfaces where slip is large are uncharged \cite{Squires2008}. Recent experiments have shown however that electro-osmosis could be amplified on superhydrophobic surfaces by polarizing the liquid-vapor interfaces \cite{Dehe2020}.

\subsection{Diffusio-osmosis: flows induced by solute gradients}\label{sec:do}

In the prospect of energy harvesting or desalination processes, solute gradients are a key aspect to consider. Indeed, the amount of fresh and salty water available naturally on earth, or the concentrated industrial dusts, could be considered as an available source of energy in large amount \cite{logan2012,Siria2017,Marbach2019}.

Here we will derive expressions for the diffusio-osmotic response coefficient. The corresponding response matrix writes: 
\begin{equation}
 \begin{pmatrix}
 q\\
 \delta j_n
 \end{pmatrix}
 = 
 \begin{pmatrix}
 \cdot & M_\text{do} \\
 M_\text{do} & \cdot
 \end{pmatrix}
 \begin{pmatrix}
 f_\text{p} = -\nabla p\\
 -\nabla \mu = -\kt\ \nabla n_0/n_0
 \end{pmatrix} , 
\end{equation}
with $q$ the flow rate density (average fluid velocity), $\delta j_n$ the excess solute flux density, which will be defined properly later, $f_\text{p} = -\nabla p$ the pressure gradient, and $-\nabla \mu = -\kt\ \nabla n_0/n_0$ the chemical potential gradient related to the gradient of solute concentration in bulk. 

In this section, we will consider both the direct diffusio-osmotic response, $q = M_\text{do} (-\nabla \mu)$, and the reciprocal streaming excess solute flux response, $\delta j_n = M_\text{do} f_\text{p}$, to obtain an expression for $M_\text{do}$. We will first consider the case of a neutral solute, and then turn to salts. 

\subsubsection{Neutral solute}

\paragraph{\textbf{Diffusio-osmotic flow}}

Let us first consider the flow induced by a gradient of solute concentration, in the absence of pressure gradient.
We consider a planar wall located at $z=0$, and a solution of B in A in the $z>0$ region. 
A solute gradient is applied far from the wall along the $x$ direction. All quantities depend on $x$, and we denote $\nabla X = \partial_x X$. 
We denote $n_\text{A}(z)$ and $n_\text{B}(z)$ the densities of A and B, and $n_\text{A}^\text{b}$ and $n_\text{B}^\text{b}$ their bulk values far from the interface. 

We will use local thermodynamics to compute the force density driving the flow in the interfacial layer~\cite{Liu2018b,Ramirez-Hinestrosa2021}. 
In the A+B mixture, the Gibbs-Duhem relation can be written $\mathrm{d}p = n_\text{A} \mathrm{d}\mu_\text{A} + n_\text{B} \mathrm{d}\mu_\text{B}$, where $p$ is the pressure, and $\mu_\text{A}$ and $\mu_\text{B}$ the chemical potential of A and B particles, respectively. 

A concentration gradient of component $i=\text{A},\text{B}$ along $x$ will lead to a chemical potential gradient $\nabla \mu_i$. 
The chemical potential for component $i$ is given by:
\begin{equation}
    \mu_i = \mu_i^0 + \kt \ln n_i^\text{b} + \mu_i^\text{exc}, 
\end{equation}
where $\mu_i^0$ denotes a (constant) reference value and $\mu_i^\text{exc}$ denotes the excess chemical potential due to intermolecular interactions.
Because the bulk solutions are ideal, $\mu_i^\text{exc}$ does not depend on the concentration of B, and the chemical potential gradients are constant along the $z$ direction; in particular, for B it writes: 
\begin{equation}
    -\nabla \mu_\text{B} 
    = -\frac{\partial \mu_\text{B}}{\partial n_\text{B}^\text{b}} \times \nabla n_\text{B}^\text{b} 
    = -\kt \frac{\nabla n_\text{B}^\text{b}}{n_\text{B}^\text{b}} 
\end{equation}

As the pressure is constant in the bulk, the Gibbs–Duhem relation reduces to $0 = n_\text{A}^\text{b} \nabla \mu_\text{A} + n_\text{B}^\text{b} \nabla \mu_\text{B}$, so that $\nabla \mu_\text{A} = -n_\text{B}^\text{b} \nabla \mu_\text{B} / n_\text{A}^\text{b}$. At a distance $z$ from the surface, a pressure gradient remains, giving a force density: 
\begin{align}
    f(z) &= -\nabla p (z) = n_\text{B}(z) (-\nabla \mu_\text{B}) + n_\text{A}(z) (-\nabla \mu_\text{A})\\
  &= (-\nabla \mu_\text{B}) \left( n_\text{B}(z) - n_\text{B}^\text{b} \frac{n_\text{A}(z)}{n_\text{A}^\text{b}}\right) 
  = -\kt \frac{\nabla n_\text{B}^\text{b}}{n_\text{B}^\text{b}} \left( n_\text{B}(z) - n_\text{B}^\text{b} \frac{n_\text{A}(z)}{n_\text{A}^\text{b}}\right). 
  \label{eq:fDO}
\end{align}

One can then implement the force density $f(z)$ in Eqs.~\eqref{eq:vosm_general} and \eqref{eq:vosm_general_stagnant} to obtain the osmotic velocity far from the wall $v^\infty_\text{osm}$. For thin interaction layers, the average flow velocity $q$ identifies with $v^\infty_\text{osm}$, and the corresponding diffusio-osmotic coefficient $M_\text{do}$ writes, on a slipping wall: 
\begin{equation}\label{eq:Mdo}
    M_\text{do} = \frac{v^\infty_\text{osm}}{-\kt \,\nabla n_\text{B}^\text{b}/n_\text{B}^\text{b}} = \frac{1}{\eta} \int_0^{\infty} \left( z + b \right) \left( n_\text{B}(z) - n_\text{B}^\text{b} \frac{n_\text{A}(z)}{n_\text{A}^\text{b}} \right) \,\dd z, 
\end{equation}
and in the stagnant layer case: 
\begin{equation}\label{eq:Mdo_noslip}
    M_\text{do} = \frac{1}{\eta} \int_{\zs}^{\infty} \left( z - \zs \right) \left( n_\text{B}(z) - n_\text{B}^\text{b} \frac{n_\text{A}(z)}{n_\text{A}^\text{b}} \right) \,\dd z.  
\end{equation}

\paragraph{\textbf{Reciprocal effect: streaming excess solute flux}}

According to Onsager reciprocal relations, the expressions of the diffusio-osmotic coefficient, Eqs.~\eqref{eq:Mdo} and \eqref{eq:Mdo_noslip}, can also be derived by considering the reciprocal effect, the streaming excess solute flux, i.e. the excess solute flux generated by a pressure gradient, in the absence of solute concentration gradient.

We consider the same slab channel that was introduced in Sec.~\ref{sec:streaming_current} on the streaming current, with two parallel walls perpendicular to the $z$ axis, located at $z=0$ and $z=d$. We apply a pressure gradient along the $x$ direction. All quantities depend on $x$, and we denote $\nabla X = \partial_x X$. As in the previous section, 
we consider a solution of B in A; we denote $n_\text{A}(z)$ and $n_\text{B}(z)$ the densities of A and B, and $n_\text{A}^\text{b}$ and $n_\text{B}^\text{b}$ their "bulk" values far from the interface. 

Finally, we denote $v(z)$ the Poiseuille velocity profile induced by the constant pressure gradient in the channel, $f_\text{p} = -\nabla p$, given by Eq.~\eqref{eq:poiseuille}.

The excess solute flux through the channel is defined as the difference between the measured solute flux and the solute flux that would be advected by the Poiseuille flow if the solute density was equal to its bulk value everywhere in the channel\footnote{{Note that this definition of the excess solute flux differs from the general expression derived for a liquid mixture in section~\ref{sec:irreversible_thermodynamics}. This is because the general expression ignores interactions between the liquid and the walls, which can generate an excess of solute density in the liquid.}}: 
\begin{equation}\label{eq:deltaJB}
    \delta J_\text{B} = J_\text{B} - n_\text{B}^\text{b} Q = 2w \int_0^{d/2} n_\text{B}(z) v(z) \,\dd z - n_\text{B}^\text{b} Q, 
\end{equation}
where $w$ is the channel width, and $Q$ is the solvent flow rate, 
defined as the change of solvent volume in the reservoir per unit time. Because in the reservoir, the solvent density is equal to its bulk value $n_\text{A}^\text{b}$, $Q$ can also be related to the flux of solvent particles, $Q = J_\text{A} / n_\text{A}^\text{b}$. The change in solvent volume per unit time in the reservoir is indeed given by the change in the number of solvent particles per unit time, divided by the bulk solvent density. \textit{In fine}, $Q$ writes: 
\begin{equation}\label{eq:QA}
    Q = \frac{J_\text{A}}{n_\text{A}^\text{b}} = \frac{2w}{n_\text{A}^\text{b}} \int_0^{d/2} n_\text{A}(z) v(z) \,\dd z.
\end{equation}
Combining Eqs.~\eqref{eq:deltaJB} and \eqref{eq:QA}, one gets: 
\begin{equation}\label{eq:deltaJB1}
    \delta J_\text{B} = 2w \int_0^{d/2} v(z) \left( n_\text{B}(z) - n_\text{B}^\text{b} \frac{n_\text{A}(z)}{n_\text{A}^\text{b}} \right) \,\dd z. 
\end{equation}
If the densities of A and B only differ from their bulk values in a thin region close to the wall where $z \ll d/2$, then the integrand in Eq.~\eqref{eq:deltaJB1} will only differ from zero in this same region. 
In that case, one can linearize the velocity profile in the integral, Eq.~\eqref{eq:vPois_lin},      
and one can extend the upper boundary of the integral to infinity: 
\begin{equation}\label{eq:deltaJB2}
    \delta J_\text{B} = \frac{w d f_\text{p}}{\eta}
    \int_0^{\infty} \left( z + b \right) \left( n_\text{B}(z) - n_\text{B}^\text{b} \frac{n_\text{A}(z)}{n_\text{A}^\text{b}} \right) \,\dd z. 
\end{equation}
The excess solute flux per unit area is then given by $\delta j_\mathrm{B} = \delta J_\mathrm{B}/(w d)$. Finally, the diffusio-osmotic response coefficient is obtained from $M_\text{do} = \delta j_\mathrm{B} / f_\text{p}$, and one recovers Eq.~\eqref{eq:Mdo}.  
The same approach can be followed to obtain the expression of the diffusio-osmotic response coefficient for a stagnant layer situation, leading to Eq.~\eqref{eq:Mdo_noslip}.

\subsubsection{Salts}

The expression of the diffusio-osmotic coefficient can easily be generalized to multiple solutes, and in particular to salts. 
Let's consider for instance the simple case of a monovalent aqueous electrolyte. We denote $n_\text{w}(z)$ the water density and $n_\text{w}^\text{b}$ its bulk value, $n_+(z)$, $n_-(z)$ the cation and anion densities and $n_0$ their bulk value. As detailed in the supporting information of Ref.~\cite{Joly2021}, the diffusio-osmotic coefficient then writes: 
\begin{equation}\label{eq:Mdo_salt}
    M_\text{do} = \frac{1}{\eta} \int_0^{\infty} \left( z + b \right) \left( n_+(z) + n_-(z) - 2 n_0 \frac{n_\text{w}(z)}{n_\text{w}^\text{b}} \right) \,\dd z  
\end{equation}
on a slipping surface, and: 
\begin{equation}\label{eq:Mdo_salt_noslip}
    M_\text{do} = \frac{1}{\eta} \int_{\zs}^{\infty} \left( z - \zs \right) \left( n_+(z) + n_-(z) - 2 n_0 \frac{n_\text{w}(z)}{n_\text{w}^\text{b}} \right) \,\dd z  
\end{equation}
in the presence of a stagnant layer. 

The diffusio-osmotic velocity profile and coefficient can then be related to the surface potential/charge of the wall by means of a few assumptions: 
\begin{itemize}
    \item ignoring the interaction and possible  layering of the solvent close to the wall, i.e. assuming $n_\text{w}(z) = n_\text{w}^\text{b}$;
    \item assuming $n_\pm(z)$ follow the Gouy-Chapman theory; 
\end{itemize}
one can write: 
\begin{equation}
    v_\text{do}(z) = \frac{1}{2\pi\lB \eta} \left\{ \ln \left[ \frac{
    1 - \left(\gamma \mathrm{e}^{-z/\debye} \right)^2
    }{1-\gamma^2}
    \right] + \frac{b}{\lGC} |\gamma| \right\}  
    \times \left(-\kt \frac{\nabla n_0}{n_0} \right) 
\end{equation}
and
\begin{equation}
    M_\text{do} = \frac{v_\text{do}(\infty)}{-\kt \,\nabla n_0/n_0} = \frac{1}{2\pi\lB \eta} \left\{ -\ln \left( 1 - \gamma^2\right) + \frac{b}{\lGC} |\gamma| \right\}, 
\end{equation}
with $\gamma$ given by Eq.~\eqref{eq:gamma_PB}. In the stagnant layer situation, this equation applies, taking $b=0$ and replacing $\phis$ by the reduced potential at the shear plane position $\zs$, where the velocity vanishes.

Note that here we only described the intrinsic diffusio-osmotic response, i.e. the coefficient in the response matrix. However, an additional flow can be generated under a salt concentration gradient, when cations and anions have a different diffusion coefficient, $D_+ \neq D_-$, and when the channel boundary conditions impose that there is no electrical current along the flow direction in the bulk liquid \cite{Anderson1989,Lee2014b}. 
Indeed, in that case a so-called diffusion electric field $E_0$ appears to avoid charge separation, which one can compute by writing that the bulk electric current vanishes: 
\begin{equation}
    j_e = q_e (j_+ - j_-) = 0, \text{  with   }
    j_\pm = -D_\pm \left( \nabla n_0 \mp \frac{q_e n_0}{\kt} E_0 \right)
    \Rightarrow E_0 = \delta \frac{\kt}{q_e} \frac{\nabla n_0}{n_0}, 
\end{equation}
with $\delta = (D_+ - D_-)/(D_+ + D_-)$. 
The diffusion electric field creates an electro-osmotic flow, which adds to the diffusio-osmotic flow, resulting in a total osmotic velocity far from the wall: 
\begin{align}
    v^\infty_\text{osm} &= M_\text{do} \times \left( -\kt \frac{\nabla n_0}{n_0} \right) + M_\text{eo} \times E_0\\
    v^\infty_\text{osm} &= \frac{1}{4\pi\lB \eta} \left\{ -2\ln \left( 1 - \gamma^2\right) + \frac{2b}{\lGC} |\gamma|  +  \delta \phis \left( 1+ \frac{b}{\lambda_\text{eff}} \right)  \right\} \times \left( -\kt \frac{\nabla n_0}{n_0} \right).  
\end{align}
In the Debye-H\"uckel limit, this expression simplifies into: 
\begin{equation}
    v^\infty_\text{osm} = \frac{\eps \Vs}{\eta} \left\{ \frac{\Vs}{8\kt} \left( 1 + \frac{2b}{\debye} \right) + \frac{\delta}{q_e} \left( 1+ \frac{b}{\debye} \right)  \right\} \times \left( -\kt \frac{\nabla n_0}{n_0} \right).  
\end{equation}
This last expression shows that the intrinsic diffusio-osmosis will dominate over the diffusion electric field induced electro-osmosis when the reduced surface potential is larger than $\delta$, and reciprocally.

\subsection{Thermo-osmosis: flows induced by thermal gradients} \label{sec:to}

Thermal gradients along liquid-solid interfaces also generate a flow, called thermo-osmotic flow \cite{Anderson1989}. The corresponding response matrix writes: 
\begin{equation}
    \mqty( q \\ \delta j_\mathrm{h}) = \mqty( \cdot & M_\tosm \\ M_\tosm & \cdot) \mqty( f_\text{p} = - \nabla p \\ -\frac{\nabla T}{T}),
    \label{eq:TO_responsematrix}
\end{equation}
where $q$ is the flow rate density (i.e. the average flow velocity),  $\delta j_\mathrm{h}$ is the heat flux density, $p$ the pressure and $T$ the temperature. 
Throughout this section, we will only consider walls with slip, but the formulas can easily be generalized to the stagnant layer case. 
Due to Onsager's reciprocal relations, there are two different paths to compute the thermo-osmotic response coefficient $M_\tosm$: the thermo-osmotic route ($q = M_\tosm ( -\nabla T/T)$), and the mechanocaloric route ($\delta j_\mathrm{h}= M_\tosm f_\text{p}$); as sketched in Fig.~\ref{fig:osmosis}. 

\subsubsection*{Thermo-osmotic flow}

First we will follow the thermo-osmotic route, $q = M_\tosm (-\nabla T / T)$. In this case, the force density driving the flow will be the thermodynamic force, $f=-T \nabla (\frac{\mu}{T})$. Taking into account the Gibbs-Helmholtz equation, $\diff{\mu/T}{T} = -\frac{\delta h}{T^2}$, where $\delta h$ is the density of enthalpy excess \cite{Bregulla2016}, we obtain that
\begin{equation}
    f(z) = -\delta h(z) \frac{\nabla T}{T}.
\end{equation}
Substituting $f(z)$ in Eq.~\eqref{eq:vosm_general} we obtain:
\begin{equation}
    v_\mathrm{osm}^\infty = - \frac{\nabla T / T}{\eta }\int_0^\infty \delta h(z) \, (z+ b) \, \dd z. 
\end{equation}
When the interaction layers are thin as compared to the channel size, the average velocity $q$ identifies with $v_\mathrm{osm}^\infty$, 
so that $M_\tosm = v_\mathrm{osm}^\infty/(-\nabla T / T)$ writes: 
\begin{equation}
    M_\tosm = \frac{1}{\eta} \int_0^\infty \delta h(z) \, (z+b) \, \dd z.
    \label{eq:Mto_gal}
\end{equation}

\subsubsection*{Mechanocaloric effect}

Equation~\eqref{eq:Mto_gal} can be also obtained, following Onsager reciprocal relations, through the mechanocaloric route, considering the heat flux generated by a pressure gradient, in the absence of temperature gradient. 
We consider the same slab channel that was introduced in Sec.~\ref{sec:streaming_current} on the streaming current, with two parallel walls perpendicular to the $z$ axis, located at $z=0$ and $z=d$. We apply a pressure gradient $f_\text{p} = -\nabla p$ along the $x$ direction, which creates a Poiseuille velocity profile $v(z)$ given by Eq.~\eqref{eq:poiseuille}. 
The average heat flux density through the channel can be written (by symmetry, one can integrate only over the bottom half of the channel): 
\begin{equation}
    \delta j_\mathrm{h} = \frac{2}{d} \int_0^{d/2} \delta h(z) \, v(z) \, \dd z.
\label{eq:heatflux_gal}
\end{equation}
{Note that we use here a definition of the heat flux, $\delta j_\mathrm{h}(z) = \delta h(z) \, v(z)$, different from the general expression derived for a liquid mixture in section~\ref{sec:irreversible_thermodynamics}, $\delta j_\mathrm{h} = \sum_n h_n m_n \rho_n \, (v_n - v)$. This is because the general expression ignores interactions between the liquid and the walls, which can generate an excess of enthalpy in the liquid.} 

If $\delta h(z)$ only differs from zero in a thin region close to the wall where $z \ll d/2$, then the integrand in Eq.~\eqref{eq:heatflux_gal} will only differ from zero in this same region. 
In that case, one can linearize the velocity profile in the integral, see Eq.~\eqref{eq:vPois_lin},      
and extend the upper boundary of the integral to infinity: 
\begin{equation}
    \delta j_\mathrm{h} = \frac{f_\text{p}}{\eta} \int_0^\infty \delta h (z) \, (z+b) \, \dd z.
\end{equation}
Taking into account from Eq.~\eqref{eq:TO_responsematrix} that $\delta j_\mathrm{h} = M_\tosm \, f_\text{p}$, one recovers the expression of $M_\tosm$ given by Eq.~\eqref{eq:Mto_gal}.

\subsubsection*{Aqueous electrolytes}

A fundamental quantity in Eq.~\eqref{eq:Mto_gal} is the enthalpy excess density $\delta h$. Here we introduce some general concepts related to $\delta h$ such as its classical description \cite{Derjaguin1941,Derjaguin1987}, given only by the ionic electrostatic interactions, together with some additional contributions that play a role in the enthalpy of an aqueous electrolyte.

In the case of aqueous electrolytes, originally, Derjaguin et al. developed a model for the thermo-osmotic coefficient \cite{Derjaguin1941,Derjaguin1987} as in Eq.~\eqref{eq:Mto_gal} without the slip term, and only considering the electrostatic enthalpy of the ions $\delta h_\mathrm{el}(z) = \rhoe(z) V(z) + p(z)$. By taking into account Eq.~\eqref{eq:Poisson} and considering mechanical equilibrium along the $z$ direction $\qty( \dv{p}{z} = - \rhoe \dv{V}{z})$, one obtains an expression of $\delta h_\mathrm{el}$ as a function of the electric potential:
\begin{equation}
    \delta h_\mathrm{el}(z) = -\eps V(z) \dv[2]{V}{z} + \frac{\eps}{2}\qty(\dv{V}{z})^2.
    \label{eq:dhel}
\end{equation}
Just focusing on this classical theory \cite{Derjaguin1987}, which only considers the electrostatic interactions between ions ($\delta h \simeq \delta h_\mathrm{el}$), substituting Eq.~\eqref{eq:dhel} in Eq.~\eqref{eq:Mto_gal}, one can solve the integral analytically in the slip situation, obtaining the electrostatic constribution to the thermo-osmotic response as a function of the ratio $x = \debye/\lGC$:
\begin{equation}
    M_\tosm^\mathrm{el} = \frac{1}{2 \pi \lB \eta \beta} \qty{ -3 \ln(1-\gamma^2) - \text{asinh}^2(x) + \frac{b}{\debye} \bigg[3 x \abs{\gamma} - 2 x \, \text{asinh}(x) \bigg]},
\end{equation}
with $\gamma$ given by Eq.~\eqref{eq:gamma_PB}.
This expression can be simplified in the Debye-H\"uckel regime, which was the one considered by Derjaguin \cite{Derjaguin1941,Derjaguin1987}, then $x \ll 1$:
\begin{equation}
    M_\tosm^\mathrm{el, DH} = - \frac{x^2}{8 \pi \lB \eta \beta} \qty( 1+2 \frac{b}{\debye}),
    \label{eq:Mtoel_DH}
\end{equation}
and thus scaling as $\Sigma^2$ in this regime.
A different scaling with $x$ is found for high surface charges, \emph{i.e.} when $x \gg 1$, when the contribution is given by the expression:
\begin{equation}
    M_\tosm^{\mathrm{el},x \gg 1} = \frac{1}{2\pi\lB \eta \beta} \qty{ 3 \ln(\frac{x}{2}) - \ln^2(2x) + \frac{b}{\debye} x \bigg[3 - 2 \ln(2x)\bigg]}.
\end{equation}
It is interesting to note that none of these expressions depend on the sign of the surface charge: $M_\tosm < 0$ independently of the range of parameters studied.

Although the model proposed by Derjaguin et al. is useful to quantitatively predict some $M_\tosm$ experimental orders of magnitude \cite{Bregulla2016}, it fails to describe the amplitude of the responses predicted in the literature \cite{Ganti2017,Fu2017,Oyarzua2017,Fu2018}, the thermo-osmotic response reported for weakly charged membranes \cite{Mengual1978}, as well as the experimental discrepancies observed in $M_\tosm$ sign \cite{Derjaguin1980,Rusconi2004,Nedev2015,Bregulla2016}. Although electrostatic ionic interactions are for sure an important ingredient controlling the thermodynamical processes of a dissolved salt in a charged channel, other interactions, which are discarded by this classical model, may also be critical to describe thermo-osmosis, such as the liquid-solid interactions (\textit{i.e.} the wetting properties), as well as the ion specificity \cite{Huang2007,Huang2008}.

Generally, the atomic enthalpy density for an element $i$ can be defined as:
\begin{equation}
    \delta h_i(z) = \qty[ \delta u_i(z) + \delta p_i(z) ] \, n_i(z),
    \label{eq:deltah_atom}
\end{equation}
where the notation $\delta$ refers to $\delta \mathcal{A}(z) = \mathcal{A}(z) - \mathcal{A}_\mathrm{bulk}$; with $\mathcal{A}_\mathrm{bulk}$ the bulk value of the physical property $\mathcal{A}$. In Eq.~\eqref{eq:deltah_atom}, $u_i$ is the energy per atom, $p_i$ the stress per atom, and $n_i$ the atomic density profile. When working at constant temperature, the kinetic energy per atom $u_{\mathrm{k},i}$ is proportional to $ k_\mathrm{B} T$ for all $z$, so $\delta u_{\mathrm{k},i} = 0$ and $\delta u_i = \delta u_{\mathrm{p},i}$ with $u_{\mathrm{p},i}$ the potential energy per atom.

Equation~\eqref{eq:deltah_atom} can be easily  extended to the case of molecular fluids as the sum of the different atomic contributions. Therefore, in the case of water $\delta h_\mathrm{wat}(z) = \delta h_\mathrm{O}(z) + \delta h_\mathrm{H}(z)$.
In this case $\delta h_\mathrm{wat}$ does not have a simple analytical form and, due to its strong dependence on the wetting properties \cite{Ganti2017,Fu2017}, it is typically computed numerically from simulations for a given wall type.

It is interesting to note that, while $M_\tosm^\mathrm{el}$ is always negative, the water contribution can change sign for different water-substrate interactions, or more specifically wetting properties \cite{Herrero2022}. Therefore, for a given salt, there can be a competition between water and electrostatic contributions, and the flow can exhibit (or not) a change of sign with $x$ depending on the amplitude of $\Sigma$. For fixed $\Sigma$, such change of sign will happen at a given $\debye$ value, implying that one can change the flow direction just by changing the salt concentration $n_0$ if the solvent term dominates. 

Note finally that a correction to $M_\tosm^\mathrm{el}$ could be accounted due to the failure of the Poisson-Boltzmann theory to capture the depletion of the ions from the wall up to a distance $d_\mathrm{l}$. Specifically, the modified electrostatic contribution writes:
\begin{equation}
    M_\tosm^{\mathrm{el}*} = \frac{1}{\eta} \int_{d_\mathrm{l}}^\infty \qty(z+b) \delta h_\mathrm{el}(z)  \,\dd z.
\end{equation}

Aside, other contributions to the total thermo-osmotic response could be considered,  as the one associated to the water molecules dipole moment in the EDL.  In this latter case, for instance, the associated density of enthalpy excess will be 
\begin{equation}
    \delta h_\mathrm{dp}(z) = - \expval{\mu}(z) \, n_\mathrm{O}(z) \, E(z),
    \label{eq:dhdp_init}
\end{equation}
with $n_\mathrm{O}$ the number density of the oxygen atoms, $E = - \dv{V}{z}$ the electrostatic field, and $\langle \mu \rangle(z) = \mu \cdot \langle \cos \theta \rangle (z)$ is the average dipole moment in the direction of the field and it is given by the expression \cite{vanderlinde2006}:
\begin{equation}
    \langle \mu \rangle = \mu \left( \coth{\alpha} - \frac{1}{\alpha} \right),
\end{equation}
where $\alpha = \beta \mu E$ and $\mu$ is the solvent's dipole moment, $\mu=1.85\,$D for water. Approximating this equation when $\mu \, E \ll 1/\beta$ (i.e. $\alpha \ll 1$) and expressing everything in terms of the reduced potential $\phi$ one finally obtains that:
\begin{equation}
    \delta h_\mathrm{dp}(z) = -\frac{1}{3 \beta} \qty(\frac{\mu}{q})^2 n_\mathrm{O}(z) \, \qty(\dv{\phi}{z})^2,
    \label{eq:dhdp}
\end{equation}
where $\phi$ is the reduced potential. One can obtain $n_\mathrm{O}$ from molecular dynamics simulations and $\phi(z)$ solving Eq.~\eqref{eq:PB} for the corresponding geometry.

\subsection{Effect of the channel geometry}

When considering transport at the nanoscale, channel geometry appears critical to optimize the transport properties and to increase energy conversion efficiency. For example, it seems tempting to decrease the channel length to reach a so-called pore geometry, to apply easily very large gradients and then to enhance the electrokinetic response. The ultimate case corresponds to pore drilled in a single graphene layer \cite{lee2014stabilization}. Other possibilities are to optimize the channel geometry, using non symmetrical devices inducing interesting non-symmetrical EK or diode-like responses, with many applicative properties \textit{e.g.} for desalination \cite{Picallo2013NanofluidicSimulations}.
Finally, the development of new technologies give access to lower and lower length scales and more and more selective channels: nanometric or even sub-nanometric sized channels. This give raise to new phenomena where our classical description  is challenged. We will briefly discuss, in the light of recent advances in the literature, these three geometrical effects on osmotic flow responses, such as channel length, channel asymmetry and channel thickness.

\subsubsection{Channel length: entrance effects}

In all theories that have been considered so far, we studied an infinitely long channel so that the velocity profile could be assumed to be fully developed. However, realistic systems have a finite length: two reservoirs are connected through a channel of length $L$ and lateral size $d$. The dimensions of the reservoirs are larger than $d$. As a consequence, entrance effects emerge and these could reduce the amplitude of osmotic response. The first known example is pressure jumps  at both entrances proportional to the average velocity in the channel, and induced by the focusing of the streamlines of the channel, which generates viscous dissipation \cite{Sampson1891,Gravelle2013}. This leads to an osmotic velocity drop. This effect can also be induced by an osmotic flow (e.g. thermo-osmotic flow) and can lead to a backflow, i.e. in the opposite direction \cite{Fu2017}. 
It has been shown recently that hydrodynamic entrance effects, in case of slippage, can be reduced with cone shape channels at a prescribed angle \cite{Gravelle2014} or a trumpet bell \cite{Belin2016}. 

Similarly, potential streamlines focusing induce a voltage drop at the entrance \cite{Hall1975, Lee2012}, limiting the gradient applied in the responsive  part of the channel, i.e. the electroosmotic response. 
Finally, when considering applications with porous membranes or many channels in parallel, it has been shown that these entrance effects can interact \cite{Gadaleta2015, Jensen2014,liot2020}, leading to 
an increase or a decrease of transport properties in the channel.

\subsubsection{Channel asymmetry}

We considered so far only symmetric parallel channels. However, some new optimization processes and functionalities can arise from different channel geometries and in particular asymmetric systems. For example, 
concerning ionic transport, nanofluidic ionic diodes have been achieved with cone shaped nanopores \cite{Vlassiouk2007}, which where even more sensitive  to pressure \cite{Jubin2018}. 
Moreover, chemically asymmetric channels have also been realised, with one side of the channel covered with cationic groups and the other side with anionic ones. In this case, ionic diode has also been experimentally observed \cite{karnik2007}, and more interestingly, a behavior as an osmotic diode has been observed  with atomistic simulations \cite{Picallo2013NanofluidicSimulations}.

\subsubsection{Channel thickness: Debye overlap}
\label{sec:debye_overlap}

Despite the complex properties, an even richer phenomenology appears in ultraconfined systems. 
Here we discuss the case of channel size $d$ smaller than the Debye length $\debye$, corresponding to overlapping EDLs. For channels larger than $\sim 1$\,nm, continuum descriptions remain valid (see Sec.~\ref{ssec:beyond_standard}), and one can still use Stokes and Poisson-Boltzmann equations to describe osmotic flows. 

Two regimes must be considered depending on the $d/\lGC$ ratio. First, when the surface charge $\Sigma$ is small enough that $d/\lGC \ll 1$, the electric potential and the ion densities are homogeneous, given by: 
\begin{equation}\label{eq:phi_overlap_smallSigma}
    \phi = \text{sgn}(\Sigma) \times \text{asinh} \left( \frac{\lDu}{d} \right) ,  
\end{equation}
\begin{equation}\label{eq:n_overlap_smallSigma}
    n_\pm = \ns \left\{ \mp \text{sgn}(\Sigma) \frac{\lDu}{d} + \sqrt{1+\left( \frac{\lDu}{d} \right)^2} \right\} ,     
\end{equation}
introducing the Dukhin length $\lDu = (|\Sigma|/q)/\ns = 4\debye^2/\lGC$, which compares the number of ions in bulk and at the interface. 

Second, when $\Sigma$ is large enough that $d/\lGC \gg 1$, co-ions are excluded from the channel and the PB equation can be solved for counter-ions only. The potential and counter-ion density then write: 
\begin{equation}\label{eq:phi_overlap_largeSigma}
    \phi(z) = -\text{sgn}(\Sigma) \times \ln\left\{\cos^2(K z)\right\}, 
\end{equation}
\begin{equation}\label{eq:n_overlap_largeSigma}
    n(z) = \dfrac{K^2}{2 \pi \lB \cos^2(K z)}, 
\end{equation}
where the inverse length $K$ is the solution of the equation: 
\begin{equation}\label{eq:Kd_vs_doverlGC}
Kd \,\tan\left(\frac{Kd}{2}\right) = \frac{d}{\lGC}.
\end{equation} 

In the overlapping Debye layer regime, one can compute the zeta potential from the EO flow rate density (i.e., the average velocity) generated by an electric field along the channel: $\overline{v_\text{eo}} = -(\eps\zeta/\eta) E$. For low surface charge, $d/\lGC \ll 1$, using Stokes equation and Eqs.~\eqref{eq:phi_overlap_smallSigma} and \eqref{eq:n_overlap_smallSigma}, one obtains: 
\begin{equation}\label{eq:zeta_overlap_smallSigma}
    \zeta = \frac{2\,\text{sgn}(\Sigma)}{\beta q} \left\{ \frac{d}{6\lGC} + \frac{b}{\lGC} \right\}. 
\end{equation}
For large surface charge, $d/\lGC \gg 1$, and using  Eqs.~\eqref{eq:phi_overlap_largeSigma} and \eqref{eq:n_overlap_largeSigma}, one obtains \cite{VanderHeyden2006}:
\begin{equation}\label{eq:zeta_overlap_largeSigma}
  \zeta = \frac{2\,\text{sgn}(\Sigma)}{\beta q} \left( \frac{1}{s} \int_0^{s} y \tan(y) \,\dd y + \frac{b}{\lGC} \right), 
\end{equation}
where $s = \frac{Kd}{2}$. We are not aware of any usable analytical solution for the integral in this expression; however, a remarkably good approximation (within $2.5\,\%$ of Eq.~\eqref{eq:zeta_overlap_largeSigma}) exists: 
\begin{equation}\label{eq:zeta_overlap_approx}
  \zeta \approx \frac{2\,\text{sgn}(\Sigma)}{\beta q} \left\{ \ln \left( 1 + \frac{d}{6\lGC} \right) + \frac{b}{\lGC} \right\}. 
\end{equation}
In practice, this expression provides a good approximation of $\zeta$ for the entire range of $d/\lGC$ values. Finally, one should note that Eq.~\eqref{eq:zeta_overlap_smallSigma}, based on the common assumption that the potential is homogeneous in the channel, overestimates $\zeta$ at large surface charges, and that Eq.~\eqref{eq:zeta_overlap_approx} is a better alternative.  

Similarly, the other response coefficients in Onsager matrix can be computed in the Debye overlap regime, based on Stokes equation, and using 
Eqs.~\eqref{eq:phi_overlap_smallSigma} and \eqref{eq:n_overlap_smallSigma} for low surface charge, when $d/\lGC \ll 1$, 
or Eqs.~\eqref{eq:phi_overlap_largeSigma} and \eqref{eq:n_overlap_largeSigma} for large surface charge, when $d/\lGC \gg 1$.

But these relations are at  equilibrium, 
and when considering ion transport, and in particular electro-osmotic and diffusio-osmotic responses, excluded ions tend to accumulate at the entrance of the channel and to be depleted at the output. This induces regulating electric fields and  so-called concentration polarization  effects \cite{Schoch2008,kim2007concentration}. The amplitude of this polarization of concentration is complex to determine as it depends on conditions at the entrance of the reservoir, i.e. on the possibility of ions to diffuse or to be advected outside entrance parts.  In this context, effects of entrance geometry have been extensively investigated \cite{mani2009propagation, kim2005modeling}. Recent analysis demonstrates that this effect should limit intrinsically the possibility of energy harvesting from chemical gradients with nanofluidic membranes \cite{wang2021}.

\subsection{Beyond standard models}
\label{ssec:beyond_standard}

As detailed above, the standard description of osmotic flows is based on macroscopic/continuum models. For instance, hydrodynamics is described through the continuum Stokes equation with a homogeneous viscosity, and the molecular nature of the liquid-solid interface is only accounted for effectively through the hydrodynamic boundary condition (slip or stagnant layer). 
Standard models ignore in particular structuring effects, interparticle correlations, and thermal fluctuations, which cannot always be neglected at the nanoscale. Detailed discussions on the limits of standard models can be found e.g. in Refs.~\cite{Hartkamp2018,Kavokine2021}. Here we will focus on the problems relevant to osmotic flows, and highlight a few recent developments going beyond standard models. 

Let us first start by common approximations, which can easily be lifted, still at the continuum level. 
For instance, the response coefficients derived in this chapter are typically expressed as a function of the surface charge, implicitly assumed to be constant. Yet, whether surface charge arises from the dissociation of surface groups or from the specific adsorption of charged species, the surface charge density is set by an electrochemical equilibrium at the interface, and depends on the pH and salt concentration: this phenomenon is known as charge regulation \cite{Markovich2016,Trefalt2016,Joly2021}. 
The surface charge is also assumed to be immobile, but when it results from the physisorption of charged species, the adsorbed ions can retain some mobility on the surface \cite{Grosjean2019}, which can dramatically modify the electrokinetic response \cite{Maduar2015,Mouterde2018,Silkina2019,Mangaud2021}. 

On a more fundamental level, standard continuum descriptions become questionable when the size of the interaction layer compares with the molecule size, as well as with the surface roughness. For instance, the thickness of the EDL where most of the ions are localized is given by the smallest of the Debye length and the Gouy-Chapman length, which can become subnanometric for realistic salt concentrations ($> 10^{-2}$\,M) or surface charges ($> 40$\,mC/m$^2$). 
In those regimes, molecular dynamics simulations, which provide an explicit description of the atomic structure and dynamics of the liquid-solid interface, can help refining the models using a bottom-up approach \cite{Rotenberg2013,Nagata2016a,Hartkamp2018}. 
In particular, specific interactions between the liquid molecules and the wall induce a layering of the fluid, and can affect in particular the local viscosity \cite{Hoang2012a,Bonthuis2013,Rezaei2021}, dielectric permittivity \cite{Bonthuis2013,Rezaei2021}, and ionic mobility \cite{Siboulet2017}; the Poisson-Boltzmann model of the ion distribution should also be modified to account for specific interactions between ions and the wall \cite{Huang2007,Huang2008,Ben-Yaakov2011,Joly2021}. 

Finally, there is growing evidence that water can couple to the electronic degrees of freedom of the solid, with e.g. consequences on liquid-solid slip \cite{Kavokine2022}, electronic current induced by water flow, or the reciprocal phenomenon \cite{Ghosh2003,Rabinowitz2020}. These results call for the development of new modeling tools, able to take into account electron dynamics, and to tackle systems large enough to describe a liquid-solid interface \cite{Kavokine2021}.

\section{Producing electricity with nanofluidic systems}

As already illustrated in section~\ref{sec:eo} with the streaming current, nanofluidic systems can also produce electricity from non-electrical thermodynamic gradients. In this section, we will describe and quantify the performance of electricity production from differences of pressure, salt concentration, and temperature.

\subsection{Basic concepts: the example of streaming current}

In section~\ref{sec:eo}, we have shown that a so-called streaming current appeared under a pressure gradient, in the absence of electric field, described by the response matrix in Eq.~\eqref{eq:Meo}. Here we will quantify how this current can be harvested.

\subsubsection{Energy conversion performance: output power and efficiency}\label{sec:efficiency}

The electrokinetic response of the fluidic device can be described in terms of total flow rate $Q$ and electric current $I$ as a function of the pressure drop $\Delta p$ and voltage $\Delta V$ across the channel \cite{VanderHeyden2006}: 
\begin{align} 
\label{eq:EK_resp_conversion}
\left\{
\begin{array}{ll}
     Q &= \frac{1}{\Zch} \Delta p + \Sstr \Delta V\\
     I &= \Sstr \Delta p + \frac{1}{\Rch} \Delta V, 
\end{array}
\right.
\end{align}
where $\Zch = \left(\pdiff{Q}{\Delta p}\right)^{-1}$ is the hydrodynamic resistance of the channel, $\Rch = \left(\pdiff{I}{\Delta V}\right)^{-1}$ is the electric resistance of the channel, and $\Sstr = \pdiff{I}{\Delta p} = \pdiff{Q}{\Delta V}$ is the streaming conductance of the channel. Equation \eqref{eq:EK_resp_conversion} corresponds to a macroscopic form of the local Onsager matrix, Eq.~\eqref{eq:Meo}, introduced earlier. 

\begin{figure}
    \centering
    \includegraphics[width=0.5\textwidth]{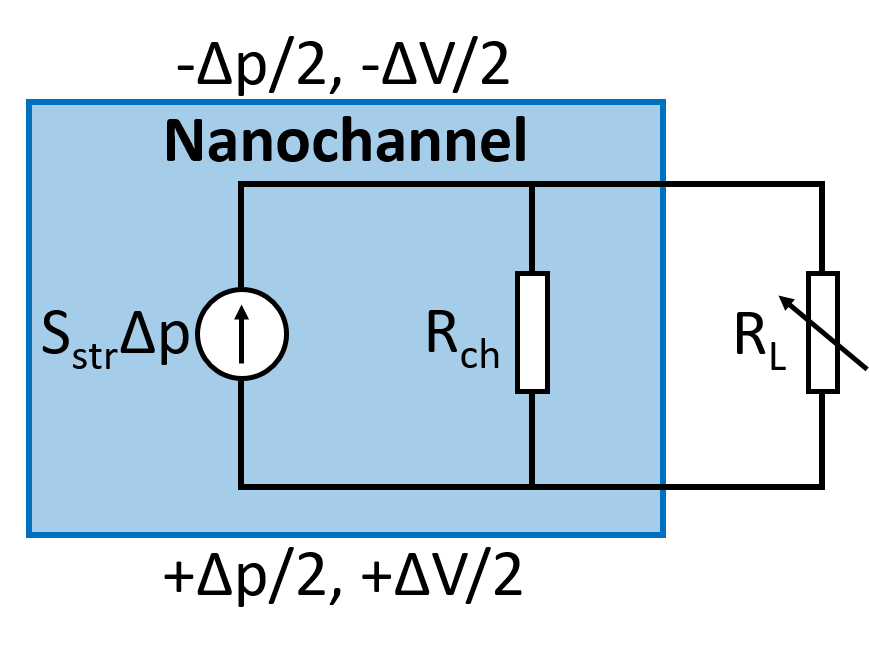}
    \caption{Equivalent electric diagram of the system used to harvest the streaming current.}
    \label{fig:equivalent_circuit}
\end{figure}

To harvest the streaming current, the nanofluidic system needs to be connected to a load. Part of the streaming current will then be distributed to the load, but another part will go back through the nanofluidic channel itself; the equivalent electric circuit can then be described by a current generator $I = \Sstr \Delta p$ with internal resistance $\Rch$, connected to an external resistor $\Rl$, see Fig.~\ref{fig:equivalent_circuit}. 

Correspondingly, the voltage across the load writes:
\begin{equation}
  \Delta V = - \frac{\Rch \Rl \Sstr \Delta p}{\Rch + \Rl}.  
\end{equation}
One can then compute the power output into the load: 
\begin{equation}
  P_\text{out} = \frac{\Delta V^2}{\Rl} = \frac{\Rch^2 \Rl \Sstr^2 \Delta p^2}{(\Rch + \Rl)^2} = \frac{\alpha \theta}{(1 + \theta)^2} \frac{\Delta p^2}{\Zch},  
\end{equation}
where $\theta = \Rl/\Rch$ and $\alpha = \Sstr^2 \Rch \Zch$. 
The maximal output power is obtained for $\theta = 1$, and reaches $P_\text{out}^\text{max} = \frac{\alpha}{4} \frac{\Delta p^2}{\Zch}$. 

Finally, one can compute the efficiency of the energy conversion as the ratio between the output power $P_\text{out}$ and the input power $P_\text{in} = Q \Delta p$: 
\begin{equation}\label{eq:ec_eff}
  \eps = \frac{P_\text{out}}{P_\text{in}} = \frac{\alpha \theta}{(1+\theta)(1+\theta-\alpha \theta)}.  
\end{equation}
The maximum efficiency $\eps_\text{max}$, is obtained for $\theta = 1/\sqrt{1-\alpha}$, and reaches: 
\begin{equation}\label{eq:ec_eff_max}
  \eps_\text{max} = \frac{\alpha}{\alpha + 2 \left(1-\alpha + \sqrt{1-\alpha} \right)}.  
\end{equation}
Therefore, $\eps_\text{max}$ varies from $0\,\%$ for $\alpha=0$ to $100\,\%$ for $\alpha=1$, which is the maximum possible value of $\alpha$ imposed by energy conservation. Note that $\eps_\text{max}$ depends non-linearly on $\alpha$, e.g. reaching only $17\,\%$ for $\alpha = 0.5$. 

Overall, to optimize the energy harvesting performance, one needs to maximize $\alpha$, i.e. find a compromise between maximizing $\Sstr$ and \emph{minimizing} the channel conductance $G_\text{ch} = 1/\Rch$. In the following, we will compute the channel conductance in order to obtain a full description of the energy conversion performance.

\paragraph*{\textbf{Electric conductivity: thin EDLs}}

We will start by considering thin EDLs, i.e. $\debye \ll d$. 
The electric current density writes: 
\begin{equation}
  j_e(z) = q_e \left[ n_+(z) v_+(z) - n_-(z) v_-(z) \right], \quad\text{with}\quad v_\pm(z) = \pm q_e \mu_\pm E_x + v_\text{eo} (z),
\end{equation}
with $q_e = Ze$, the absolute ion charge (with $Z$ the valence), $\mu_\pm$, the ion mobility, $E_x$, electric field parallel to the interface and $v_{eo}$, the electro-osmotic velocity given by equation \eqref{eq:vEO_gal}.
The electric conductivity $\sigma$ (averaged over the channel section) then writes (detailed calculations can be found in Refs.~\cite{Bocquet2009,Balme2015,Werkhoven2020}, and useful formulas in Ref.~\cite{Herrero2021}): 
\begin{align}
  \sigma &= \frac{I/(wd)}{E_x} = \frac{2}{E_x d} \int_0^{d/2} j_e(z) \,\dd z\\ 
  &= 2 q_e^2 \bar{\mu} n_0 +\frac{1}{d} \left\{ 2 q_e^2 \bar{\mu} \beta\mathcal{E} - 2q_e \Delta\mu \Sigma + \frac{q_e^2 \beta\mathcal{E}}{\pi \lB \eta} + \frac{2 b \Sigma^2}{\eta} \right\}, \label{eq:conductivity}
\end{align}
where $\bar{\mu} = \frac{\mu_+ + \mu_-}{2}$, $\Delta \mu = \frac{\mu_+ - \mu_-}{2}$, and $\mathcal{E}$ is the electrostatic energy (per unit area) of the EDL on one wall: $\mathcal{E} = \frac{\eps}{2} \int_0^{d/2} \left( \diff{V}{z} \right)^2 \,\dd z = \frac{1}{2\pi \lB \debye \beta} \left\{ -1 + \sqrt{1+(\debye/\lGC)^2} \right\}$ \cite{Herrero2021}. 
The first term in Eq.~\eqref{eq:conductivity} is the bulk ionic conductivity, and the terms between the curly brackets are surface contributions, respectively, from ionic mobility (with a first term controlled by the average mobility $\bar{\mu}$ and a second controlled by the mobility asymmetry $\Delta \mu$), and from advection by the electro-osmotic flow (with a first term related to the no-slip part of the flow, and an additional term due to slip).

The electric conductance $1/\Rch$ is then simply given by $1/\Rch = \sigma \times S/L$, with $S$ and $L$ the channel cross section and length.
One can note that, while the streaming conductance $\Sstr$ arises from the EDL and is controlled by surface charge, the electric conductance also includes a contribution from the bulk, proportional to the salt concentration and to the channel cross section. Accordingly, to maximise $\Sstr$ while keeping $1/\Rch$ as low as possible, one need to minimize the channel lateral size and the salt concentration (therefore increasing the Debye length). As a result, the channel size can become smaller than the Debye length, and one needs to compute the electric conductivity in the Debye overlap regime. 

\paragraph*{\textbf{Electric conductivity: Debye overlap}}

When the EDLs overlap, $d/\debye \ll 1$, and when the surface charge is low enough that $d/\lGC \ll 1$, the electric potential and ion densities are homogeneous, given by Eqs.~\eqref{eq:phi_overlap_smallSigma} and \eqref{eq:n_overlap_smallSigma}. The corresponding conductivity (averaged over the channel section) writes: 
\begin{equation}
    \sigma = 2 q_e^2 \bar{\mu} \sqrt{\ns^2 + \left( \frac{\Sigma}{q_e d} \right)^2} - \frac{2 q_e \Delta \mu \Sigma}{d} + \frac{\Sigma^2}{3\eta} \left( 1 + \frac{6b}{d} \right) , 
\end{equation}
where in particular $\ns$ is the salt concentration \emph{in the reservoirs}. The first two terms come from ionic mobility (with a first term controlled by the average mobility $\bar{\mu}$ and a second controlled by the mobility asymmetry $\Delta \mu$), and the third term comes from advection by the electro-osmotic flow.

At high surface charge, when $d/\lGC \gg 1$, co-ions are excluded from the channel, and the electric potential and counter-ion density are given by Eqs.~\eqref{eq:phi_overlap_largeSigma} and \eqref{eq:n_overlap_largeSigma}. The (average) electric conductivity then writes: 
\begin{equation}
  \sigma = \frac{2 q_e \mu_\text{ci} |\Sigma|}{d} + \frac{q_e^2 s (\tan s - s)}{\pi^2 \lB^2 d^2 \eta} + \frac{2 b \Sigma^2}{d \eta}, 
\end{equation}
where $s = Kd/2$, and $\mu_\text{ci} = \bar{\mu} - \text{sgn}(\Sigma) \Delta \mu$ is the counter-ion mobility; here the first term comes from ionic mobility, and the two last terms from advection by the electro-osmotic flow (with the second term, denoted $\sigma_\text{EO}$ below, related to the no-slip part of the flow, and the third to the slip contribution).  
The no-slip EO contribution, $\sigma_\text{EO}$, is only expressed implicitly as a function of surface charge, through the relation: $2s \tan s = d/\lGC$. However, a very good approximation (less than 1\,\% error everywhere) can be written: 
\begin{equation}
    \sigma_\text{EO} \approx \frac{\Sigma^2}{3\eta} \times \frac{1}{1-\sqrt{d/\lGC}/30+d/(6\lGC)}
\end{equation}
Simpler approximate expressions are obtained in the limit of low and high surface charges. For low $|\Sigma|$, i.e. when $d \ll \lGC$, 
\begin{equation}
  \sigma_\text{EO} \approx \frac{\Sigma^2}{3\eta}, 
\end{equation}
and for high surface charges, i.e. when $d \gg \lGC$, 
\begin{equation}
  \sigma_\text{EO} \approx \frac{q_e |\Sigma|}{\pi \lB d \eta}.
\end{equation}

\subsection{Harvesting blue energy: reverse electrodialysis}

We have seen in section~\ref{sec:do} that gradients of salt generated a so-called diffusio-osmotic flow. The charge in the EDL is then advected by this flow, leading to a so-called diffusio-osmotic current \cite{Siria2017}. 

\subsubsection{Salt-gradient induced current: thin EDLs}
\label{sec:Kosm_surf}

The response matrix describing the diffusio-osmotic current is: 
\begin{equation}
 \begin{pmatrix}
 j_e\\
 \delta j_n
 \end{pmatrix}
 = 
 \begin{pmatrix}
 \cdot & M_\text{en} \\
 M_\text{en} & \cdot
 \end{pmatrix}
 \begin{pmatrix}
 E=-\nabla V\\
 -\kt\ \nabla n_0/n_0
 \end{pmatrix} . 
\end{equation}
Accordingly, the response coefficient $M_\text{en}$ can be calculated in two ways \cite{Joly2021}: 
\begin{enumerate}
    \item by considering the so-called electrodialysis response, i.e. the excess solute flux density $\delta j_n$ induced by an electric field $E$, in the absence of salt gradient ($\nabla n_0 = 0$): 
    \begin{equation}
        M_\text{en} = \frac{\delta j_n}{E}
        = \frac{2}{dE} \int_{0}^{d/2} \left( n_+(z) + n_-(z) - 2 n_0 \frac{n_\text{w}(z)}{n_\text{w}^\text{b}} \right) v_\text{eo}(z) \,\dd z , 
    \end{equation}
    where 
    the expression of the electro-osmotic velocity profile $v_\text{eo}(z)$ can be found in section~\ref{sec:eo};
    \item by considering the diffusio-osmotic current, i.e. the electrical current density $j_e$ induced by a gradient of salt $\nabla n_0$, in the absence of electric field ($E=0$): 
    \begin{equation}
        M_\text{en} = \frac{j_e}{-\kt\ \nabla n_0/n_0}
        = \frac{2}{-d \kt\ \nabla n_0/n_0} \int_{0}^{d/2} \rhoe(z) v_\text{do}(z) \,\dd z ,  
    \end{equation}
    where the diffusio-osmotic velocity profile $v_\text{do}(z)$ is discussed in section~\ref{sec:do};
\end{enumerate}

Regardless of the route chosen, an analytical expression can be derived for $M_\text{en}$ in the case of thin EDLs -- i.e. $\debye\ll d$,  
by ignoring the layering of the solvent close to the wall -- i.e. assuming $n_\text{w}(z) = n_\text{w}^\text{b}$, and by assuming that $n_\pm(z)$ follow the Gouy-Chapman theory \cite{Siria2013}:
\begin{equation}
    M_\text{en} = \frac{-\Sigma}{\pi\lB\eta d} \left\{ 1 - \frac{\text{asinh}(x)}{x} + \frac{b|\gamma|}{\lGC} \right\}, 
\end{equation}
where $x=\debye/\lGC$ and $\gamma$ is given by Eq.~\eqref{eq:gamma_PB}.

Two limits can be considered.
In the Debye-H\"uckel limit at low surface charge, i.e. when $x\ll 1$, $M_\text{en}$ scales as $\Sigma^3$: 
\begin{equation}
    M_\text{en}^\text{D-H} \approx \frac{-\Sigma^3 \debye^2}{6 \eps \eta d \kt} \left( 1 + \frac{3b}{\debye} \right). 
\end{equation}
In contrast, at high surface charge, i.e. when $x\gg 1$, $M_\text{en}$ scales as $\Sigma$:
\begin{equation}
    M_\text{en}^{x\gg 1} = \frac{-\Sigma}{\pi\lB\eta d} \left( 1 + \frac{b}{\lGC} \right). 
\end{equation}

Importantly, $M_\text{en}$ scales as the inverse of the channel size $d$: better performance are obtained for smaller channels. However at some point the approximation of thin interfacial layers will fail, and one has to take the overlap of the EDLs into account, see section~\ref{sec:debye_overlap}.

\subsubsection{Bulk contribution to the current}

Contrary to osmotic flows and their reciprocal effects, which can only be generated at interfaces (see section~\ref{sec:irreversible_thermodynamics}), an electric current can be generated by a gradient of solute concentration even in bulk. In practice, such phenomenon will occur when cations and anions have different diffusion coefficients. 

Still considering a slit channel with walls at $z=0$ and $z=d$, when a gradient of salt concentration is applied along the $x$ direction, ions will move by diffusion and the resulting flux densities for cations $j_+$ and for anions $j_-$ are: 
\begin{equation}
    j_\pm = -D_\pm \nabla n_\pm = \frac{n_\pm D_\pm}{\kt} \times \left( -\kt \frac{\nabla n_\pm}{n_\pm} \right).  
\end{equation}
Because the chemical potential gradients are constant along the $z$ direction, one can compute them in bulk where $n_\pm = \ns$, so that $-\kt \nabla n_\pm / n_\pm = -\kt \nabla \ns / \ns$. 
The resulting current is $j_e = q (j_+ - j_-)$, and the corresponding (local) Onsager response coefficient writes: 
\begin{equation}
    M_\text{en} = \frac{j_e}{-\kt\ \nabla n_0/n_0}
        = \frac{q}{\kt} \left( n_+ D_+ - n_- D_- \right).
\end{equation} 
In bulk where $n_\pm=\ns$, 
\begin{equation}
    M_\text{en}^\text{bulk}  
        = \frac{q \ns}{\kt} \left( D_+ - D_- \right).
\end{equation} 

According to Onsager reciprocal relations, one can also compute the response coefficient by considering the solute flux generated by an electric field $E=-\nabla V$. Indeed, in that case the ion flux densities are: $j_\pm = \pm q n_\pm D_\pm /(\kt) \times (-\nabla V)$. Writing $\delta j_n = j_+ + j_-$ and computing $M_\text{en} = \delta j_n / (-\nabla V)$, one recovers the expressions above.  

The bulk contribution to the current is not necessarily negligible as compared to the surface one discussed in section~\ref{sec:Kosm_surf}. Let's consider for example the large diffusio-osmotic currents reported by Siria \textit{et al.} \cite{Siria2013} in boron nitride nanotubes. In a tube of radius $R=40$\,nm and length $L=1250$\,nm, so-called osmotic mobilities $K_\text{osm} = I_e/(\Delta \ns / \ns)$ up to $\sim 0.15$\,nA are reported, and attributed to a surface contribution. The bulk contribution to $K_\text{osm}$ is: 
\begin{equation}
    K_\text{osm}^\text{bulk} = \frac{\kt \pi R^2}{L} M_\text{en}^\text{bulk} = \frac{\pi R^2 q \ns}{L} \left( D_+ - D_- \right).
\end{equation}
One can estimate $K_\text{osm}^\text{bulk}$, in the same tube, at high salt concentration $\ns = 1$\,M, for different salts: $|K_\text{osm}^\text{bulk}| \sim 0.03$, $0.3$, and $3$\,nA for KCl, NaCl and HCl respectively 
($D_{\text{K}^+}=1.96\times 10^{-9}$\,m$^2$/s, 
$D_{\text{Na}^+}=1.33\times 10^{-9}$\,m$^2$/s,
$D_{\text{H}^+}=9.31\times 10^{-9}$\,m$^2$/s, 
$D_{\text{Cl}^-}=2.03\times 10^{-9}$\,m$^2$/s). 
Of course, one needs to go to very large salt concentrations (or to use larger channels) to have such high bulk contributions, which decreases the electrical resistance of the channel, with consequences on the energy conversion performance, see next section.

\subsubsection{Energy conversion performance: output power and efficiency}
As for the streaming current situation defined in section~\ref{sec:efficiency}, the nanochannel can be considered  as a current generator and the output power is simply defined by $P_{\text{out}}= \Delta V \times I_\text{osm}$ \cite{Kim2010,Siria2013}.
Practically, to harvest the diffusio-osmotic current, the nanofluidic system needs to be connected to a load. One can compute the ouput power and efficiency of the energy conversion in that setup.
Here the EK response of the fluidic channel, with a cross-section area $A$ and an length $L$, writes: 
\begin{align} 
\label{eq:EK_resp_conversion_DO}
\left\{
\begin{array}{ll}
     \delta J_n &= \frac{1}{\Zch} (\kt \Delta n_0 / n_0) + \Sdoc \Delta V\\
     I &= \Sdoc (\kt \Delta n_0 / n_0) + \frac{1}{\Rch} \Delta V, 
\end{array}
\right.
\end{align}
where $\Rch$ is again the resistance of the channel, $1/\Zch$ 
quantifies ion diffusivity through the channel, and $\Sdoc = A M_\text{en} / L$. 

Expressions analog to the streaming current case can then be derived,  for the power output: 
\begin{equation}
  P_\text{out} = \frac{\alpha \theta}{(1 + \theta)^2} \frac{(\kt \Delta n_0 / n_0)^2}{\Zch} , 
\end{equation}
where $\theta = \Rl/\Rch$ and $\alpha = \Sdoc^2 \Rch \Zch$.
The maximum output power is obtained for $\theta=1$ \cite{Kim2010} and reads
 \begin{equation}
 P_\text{out}^\text{max} = \frac{\alpha}{4} \frac{(\kt \Delta n_0 / n_0)^2}{\Zch}.
 \end{equation}
The maximum efficiency, defined by \eqref{eq:ec_eff_max} can be calculated knowing that the input chemical power is  $P_\text{in} = \delta J_n \times (\kt \Delta n_0 / n_0)$.

\subsection{Harvesting waste heat: thermoelectricity}

\subsubsection{Thermal gradient induced current: thin interaction layers} 

One can also generate an electric current by applying a temperature gradient, and the process is known as thermoelectricity \cite{Dietzel2016,Fu2019}. The corresponding response matrix writes: 
\begin{equation}
      \mqty( j_e \\ \delta j_h) = \mqty( \sigma & M_\tel \\ M_\tel & k T) \mqty( E=-\nabla V \\ -\frac{\nabla T}{T}),
    \label{eq:TE_responsematrix}
\end{equation}
where $\delta j_\mathrm{h}$ is the heat flux density, $j_\mathrm{e}$ the electric flux density, $E = - \nabla V$ the external electric field parallel to the interface, $T$ the temperature, $\sigma$ the electrical conductivity, and $k$ the thermal conductivity of the channel. In order to compute the thermoelectric response coefficient $M_\tel$, we will follow the relation $\delta j_\mathrm{h}=M_\tel \, E$, although, due to Onsager's reciprocal relations, the same result is obtained from the relation $j_\mathrm{e} = - M_\tel \nabla T / T$. Analogously to the thermo-osmotic response computation through the mechanocaloric route in the limit of thin interactions layers, the average heat flux density in a channel of height $d$ is given by Eq.~\eqref{eq:heatflux_gal}, although in this case the velocity field we should consider is the one induced by an electric field, also known as electro-osmotic velocity profile, which is given by Eq.~\eqref{eq:vEO_ch}. 
One finally obtains that 
\begin{equation}
    M_\tel = \frac{q_e}{2 \pi \lB d \eta} \int_0^\infty \delta h(z) \, \left[ \phi(z) - \phis + b \eval{\dv{\phi}{z}}_{z=0} \right]  \,\dd z ,
    \label{eq:M_tel}
\end{equation}
where, analogously to the thermo-osmotic response situation, one can consider that the main contributions to $\delta h(z)$ are the same as discussed in Sec.~\ref{sec:to} \cite{Fu2019,Herrero2022}.

We can compute the classical electrostatic contribution to the thermoelectric response in the slip situation. In this case, $\eval{\dv{\phi}{z}}_{z=0} = - \frac{2 \sgn(\Sigma)}{\lGC}$ (see Ref.~\cite{Herrero2021}), and the integral in Eq.~\eqref{eq:M_tel} can be performed analytically, giving:
\begin{equation}
\begin{split}
    M_\tel^\mathrm{el} = - \frac{q}{2\pi^2\lB^2d \eta \beta } \frac{\sgn(\Sigma) x}{\debye} & \Bigg\{ 5 \qty[1 - \frac{\text{asinh}(x)}{x}] - 2 \abs{\gamma}\text{asinh}(x) \\
    & + \frac{b}{\debye}  \bigg[ 3 \abs{\gamma} x - 2 x \text{asinh}(x) \bigg] \Bigg\};
\end{split}
\end{equation}
with $x=\debye/\lGC$ and $\gamma$ defined in Eq.~\eqref{eq:gamma_PB}. As for thermo-osmosis, it is interesting to obtain the simplified expression corresponding to low and high surface charge limits. Therefore, in the Debye-H\"uckel regime (when $x = \debye/\lGC \ll 1$):
\begin{equation}
    M_\tel^\mathrm{el,DH} = \frac{q}{12\pi^2\lB^2d \eta \beta} \frac{\sgn(\Sigma) x^3}{\debye} \qty( 1+3\frac{b}{\debye} );
\end{equation}
which, instead of scaling as $\Sigma^2$ like the thermo-osmotic response Eq.~\eqref{eq:Mtoel_DH}, scales as $\Sigma^3$. 
In the Gouy-Chapman regime corresponding to high surface charges, \emph{i.e.} when $x\gg 1$, the scaling changes:
\begin{equation}
    M_\tel^{\mathrm{el},x \gg 1} = -\frac{q}{2 \pi^2 \lB^2 d \eta \beta} \frac{\sgn(\Sigma) x}{\debye} \qty{  5 - 2 \ln(2x)  + \frac{b}{\debye} x \left[ 3 - 2 \ln(2x) \right] }.
\end{equation}

\subsubsection{Bulk contribution to the current}

As discussed for currents induced by salt gradients, an electric current can be generated by a temperature gradient even in the absence of interfaces. In practice this is due to the so-called thermal diffusion of the ions. The interested reader can find more details in Refs.~\cite{deGrootMazur,hafskjold1993molecular,Galliero2006,Artola2008,Wurger2010,Hannaoui2013}.

\subsubsection{Energy conversion performance: Seebeck coefficient, thermoelectric figure of merit}

Thermoelectricity, in general, encompasses three effects:
\begin{itemize}
\item The Seebeck effect, which was first discovered by the Estonian physicist T. Seebeck in 1821, and corresponds to the production of an electric current across the junction between two conductors having different temperatures;
\item The Peltier effect, which was discovered by J. Peltier in 1834, and is the reverse of the Seebeck effect: Peltier observed that heat is exchanged across the junction between two conductors crossed by an electric current; note that Peltier effect may produce either heating or cooling, depending on the direction of the electric current; 
\item the Thomson-Joule effect, which appears when a temperature gradient is applied to a conductor under non-isothermal conditions; in this situation, the heat flow exchanged with the surrounding medium is the sum of three contributions: thermal conduction, Joule effect and the Thomson effect. 
\end{itemize}

Let us briefly discuss how to quantify the performance of thermoelectric devices, which 
rely either on the Seebeck effect (power generation) or on the Peltier effect (thermoelectric refrigeration). 
To introduce the figure of merit, let us first concentrate on a solid state power generator, which is supposed to be, for the sake of simplicity, a $1$D conductor.
The conductor is in contact with a hot source at $x=0$ (temperature $T_H$) and with a cold source at $x=L$ (temperature $T_C$). Under steady state conditions, and if we suppose the transport properties to be weakly dependent of temperature, the temperature profile $T(x)$ is given by the solution of:
\begin{equation}
\label{eq:steady_state_temperature_profile}
    0=k \frac{d^2 T}{dx^2}+\frac{J_e^2}{\sigma}. 
\end{equation}
The electric current $J_e$ is uniform in the steady state, so that the solution of Eq.~\eqref{eq:steady_state_temperature_profile} is: $T(x)=T_H+\frac{x}{L}(T_c-T_H)+\frac{J_e^2}{2\sigma k}x(L-x)$. The heat flux current is:
\begin{equation}
   \label{eq:heat_flux_1D}
    J_Q=T(x)S J_e -k\frac{dT}{dx}. 
\end{equation}
The power generator efficiency $\eta$ is defined as the ratio of the output power over the heat supplied at the hot side: 
 \begin{equation}
     \eta = \frac{J_e \Delta V}{J_Q}
 \end{equation}
Using the expression of the heat flux eq.~\ref{eq:heat_flux_1D}, and of the voltage difference 
$\Delta V=S \Delta T+J_e L/\sigma$, the efficiency takes the form:
\begin{equation}
\eta(J_e)=\frac{J_e (S \Delta T + J_eL/\sigma)}{SJ T_H - \frac{k(T_H-T_C)}{L}+\frac{J_e^2 L}{2 \sigma}}. 
\end{equation}
The maximal efficiency is obtained for the current $J_{\rm max}$ maximizing $\eta(J)$: 
$\frac{d \eta}{d J_e}=0$. The corresponding efficiency can be cast in the form:
\begin{equation}
\label{eq:maximal_efficiency}
    \eta(J_{\rm max})=\frac{(T_H-T_C)(\sqrt{ZT_M +1}-1)}{T_H\sqrt{Z T_M+1} + T_C}, 
\end{equation}
where $T_M=(T_H+T_C)/2$ is the average temperature of the conductor, and we have introduced the figure of merit $Z$:
\begin{equation}
    Z=\frac{S^2 \sigma}{k}
\end{equation}
Because $Z$ has the units of the inverse of a temperature, it is convenient to introduce the dimensionless figure of merit $ZT$. A perfect thermoelectric material has a $ZT \gg 1$, which gives a maximal efficiency $\eta_C = \frac{(T_H-T_C)}{T_H}$, which is nothing else than the ideal Carnot efficiency. Hence, a good thermoelectric material should have a high Seebeck coefficient, a high electric conductivity $\sigma$ and a low thermal conductivity $k$. 
The best $ZT$ materials are doped semiconductors. Metals have high electric conductivity, but 
moderate Seebeck coefficient and high thermal conductivity. In semiconductors, the thermal conductivity is dominated by lattice vibrations (phonons) and, thus, can be reduced either by alloying or nanostructuring the material. Good thermoelectric materials, with a $ZT$ close to $1$ at room temperature are Bi$_2$Te$_3$, Se$_2$Te$_3$ and PbTe.

Coming back to nanofluidic systems, 
the Seebeck coefficient is defined as $S_e = -\nabla V / \nabla T$ when $j_e = 0$. It then results from Eq.~\eqref{eq:TE_responsematrix} that $S_e = M_{te} /(\sigma T)$. 
$ZT$ is expressed as a function of the Seebeck coefficient $S_e$, the thermal conductivity, the electric conductivity and the temperature: $ZT = \sigma S_e^2 T/ k$. The figure of merit can equivalently be expressed as a function of the thermoelectric coefficient $M_{\rm te}$: $ZT = M_{\rm te}^2 / (\sigma k T)$. 
Therefore, in contrast with solid-state thermoelectrics, for a nanofluidic system with a given $M_{\rm te}$, $ZT$ is optimized for channels with the lowest thermal conductivity \emph{and electrical conductivity}. The thermal conductivity of the nanofluidic system depends on the liquid, which provides little flexibility, and by the material of the surfaces, for which a low thermal conductivity material should be preferred. To minimize the electrical conductivity of the liquid, one should work at the lowest possible salt concentration, and with channels as small as possible. In that limit however, the EDLs will overlap and the expressions for $M_{\rm te}$ obtained in the thin EDL limit should be extended.

\section{Conclusions and perspectives}

When water meets a solid surface, a variety of mechanisms generate a surface charge, and ions in the liquid reorganize to form a diffuse neutralizing layer, the electrical double layer (EDL). 
Thermodynamic gradients and fluxes of different nature (hydrodynamical, electrical, chemical, thermal) can be coupled through the EDL. Such couplings are referred to as electrokinetic effects (EK), and can be used for energy conversion/harvesting using nanofluidic membranes. 
This chapter presented how the amplitude of EK effects could be predicted based on a microscopic description of the EDL, focusing on osmotic flows (flows generated at a liquid-solid interface by a non-hydrodynamic actuation), and on electric current generation from non-electrical actuation. 
Within the state-of-the-art models presented in this chapter, one can identify a number of key parameters to optimize the energy conversion efficiency. 
At the scale of the EDL, the response can in particular be boosted by hydrodynamic slip, appearing on special surfaces with very low friction. 
At the scale of the nanofluidic system, an optimum has to be found between maximizing the EK response, while minimizing the channel electrical conductance. 
Because EK effects originate at the surfaces of the channel in the EDL, while the electrical conductance arises both from bulk and surface ions, optimal performances are reached by reducing the salt concentration (hence increasing the thickness of the EDL), and by reducing the channel size. This results in an overlap of the EDL, where alternative analytical descriptions are required, which were briefly introduced in this chapter. 
However, a more radical change of paradigm is needed when the channel dimensions become comparable to the molecule size. Here continuum descriptions fail, and new and exciting behaviors can appear, which can certainly help improving the performance of nanofluidic systems, and extend the range of their applications. It is therefore crucial to explore further the new phenomena appearing at the molecular scale. To that aim, the traditional toolbox of continuum, mean field models of hydrodynamics and electrostatics will not be enough, and will need to be complemented with statistical physics \cite{Hartkamp2018,Kavokine2021}. At extreme confinements, the quantum nature of the liquid-solid interface will even need to be taken into account \cite{Kavokine2021}. For sure, the water-solid interface has not revealed all its secrets.

\begin{acknowledgement}
SM acknowledges exchange with S. Pailh\`es and J.~F. Robillard.  
LJ acknowledges interesting exchanges with E. Trizac. 
The authors are grateful to C. Ybert, D. Frenkel and G. Galliero for their feedback. 
%If you want to include acknowledgments of assistance and the like at the end of an individual chapter please use the \verb|acknowledgement| environment -- it will automatically be rendered in line with the preferred layout.
\end{acknowledgement}

%\bibliographystyle{alpha}
%\bibliographystyle{spphys}
%\bibliography{laurent,cecilia}

\end{document}